\documentclass{article} 
\usepackage{iclr2024_conference,times}

\usepackage{amsmath,amsfonts,bm}

\def\eqref#1{equation~\ref{#1}}

\def\1{\bm{1}}

\DeclareMathAlphabet{\mathsfit}{\encodingdefault}{\sfdefault}{m}{sl}
\SetMathAlphabet{\mathsfit}{bold}{\encodingdefault}{\sfdefault}{bx}{n}

\usepackage{hyperref}
\usepackage{url}
\usepackage[utf8]{inputenc}
\usepackage[T1]{fontenc}
\usepackage{color,soul}
\usepackage{url}            
\usepackage{booktabs}       
\usepackage{amsfonts}       
\usepackage{nicefrac}       
\usepackage{microtype}      
\usepackage{graphicx}
\usepackage{tabularx} 
\usepackage{subcaption}
\usepackage{float}
\usepackage{booktabs}
\usepackage{wrapfig}
\usepackage{adjustbox}
\usepackage{multirow}
\usepackage[ruled]{algorithm2e}

\SetKwInput{kwInput}{\textbf{Input}}
\SetKwInput{kwInit}{\textbf{Initialization}}
\newcommand{\calM}[0]{\mathcal{M}}
\newcommand{\calO}[0]{\mathcal{O}}
\newcommand{\xhdr}[1]{{\noindent\bfseries #1}}
\newcommand{\lz}[0]{{\sc LambdaZero}~}

\title{Generative Active Learning for the Search of Small-molecule Protein Binders}

\author{Maksym Korablyov\textsuperscript{1,4, $\star$}, Cheng-Hao Liu\textsuperscript{1,4,5, $\star$}, Moksh Jain\textsuperscript{1,3, $\star$}, Almer van der Sloot\textsuperscript{1,2,3, $\star$}
\\ 
\bf Éric Jolicoeur\textsuperscript{2, $\star$}, Edward Ruediger\textsuperscript{2, $\star$}, Andrei Nica\textsuperscript{1, $\star$}, Emmanuel Bengio\textsuperscript{6},  
\\
\bf Kostiantyn Lapchevskyi\textsuperscript{1}, Daniel St-Cyr\textsuperscript{7}, Doris Alexandra Schuetz\textsuperscript{2}, Victor Ion Butoi\textsuperscript{8},  
\\
\bf Jarrid Rector-Brooks\textsuperscript{1,3,4}, Simon Blackburn\textsuperscript{1}, Leo Feng\textsuperscript{1,3}, Hadi Nekoei\textsuperscript{1,3}, Saikrishna Gottipati\textsuperscript{9}, 
\\
\bf Priyesh Vijayan\textsuperscript{1,3}, Prateek Gupta\textsuperscript{10}, Ladislav Rampášek\textsuperscript{11}, Sasikanth Avancha\textsuperscript{12}, 
\\
\bf Pierre-Luc Bacon\textsuperscript{1,3,$\ddag$}, William Hamilton\textsuperscript{1,5}, Brooks Paige\textsuperscript{13}, Sanchit Misra\textsuperscript{12}, 
\\
\bf Stanislaw Jastrzebski\textsuperscript{14}, Bharat Kaul\textsuperscript{12}, Doina Precup\textsuperscript{1,5,15,$\ddag$}, José Miguel Hernández-Lobato\textsuperscript{16}, 
\\
\bf Marwin Segler\textsuperscript{17}, Michael Bronstein\textsuperscript{10}, Anne Marinier\textsuperscript{2}, Mike Tyers\textsuperscript{2,18,19}, Yoshua Bengio\textsuperscript{1,3,$\dagger$} \\
\textsuperscript{1}Mila -- Qu\'ebec AI Institute\quad
\textsuperscript{2}IRIC, Universit\'e de Montr\'eal\quad 
\textsuperscript{3}DIRO, Universit\'e de Montr\'eal\\
\textsuperscript{4}Dreamfold\quad
\textsuperscript{5}McGill University\quad
\textsuperscript{6}Valence Labs\quad
\textsuperscript{7}X-Chem\quad
\textsuperscript{8}MIT\quad
\textsuperscript{9}AI Redefined\\
\textsuperscript{10}University of Oxford\quad
\textsuperscript{11}Isomorphic Labs\quad
\textsuperscript{12}Intel\quad
\textsuperscript{13}University College London\\
\textsuperscript{14}Molecule.one\quad
\textsuperscript{15}Google DeepMind\quad
\textsuperscript{16}University of Cambridge\quad
\textsuperscript{17}Microsoft Research\\
\textsuperscript{18}The Hospital for Sick Children Research Institute\quad
\textsuperscript{19}University of Toronto\\
\textsuperscript{$\star$}Equal contribution\quad
\textsuperscript{$\ddag$}CIFAR AI Chair\quad
\textsuperscript{$\dagger$}CIFAR Senior Fellow\\
\texttt{\{liucheng,moksh.jain,almer.van-der-sloot\}@mila.quebec}
}

\iclrfinalcopy 
\begin{document}

\maketitle

\begin{abstract}
Despite substantial progress in machine learning for scientific discovery in recent years, truly \emph{de novo} design of small molecules which exhibit a property of interest remains a significant challenge. We introduce \lz, a generative active learning approach to search for synthesizable molecules. Powered by deep reinforcement learning, \lz learns to search over the vast space of molecules to discover candidates with a desired property. We apply \lz with molecular docking to design novel small molecules that inhibit the enzyme soluble Epoxide Hydrolase 2 (sEH), while enforcing constraints on synthesizability and drug-likeliness. \lz provides an exponential speedup in terms of the number of calls to the expensive molecular docking oracle, and \lz de novo designed molecules reach docking scores that would otherwise require the virtual screening of a hundred billion molecules. Importantly, \lz discovers novel scaffolds of synthesizable, drug-like inhibitors for sEH. In \emph{in vitro} experimental validation, a series of ligands from a generated quinazoline-based scaffold were synthesized, and the lead inhibitor N-(4,6-di(pyrrolidin-1-yl)quinazolin-2-yl)-N-methylbenzamide (UM0152893) displayed sub-micromolar enzyme inhibition of sEH. 
\end{abstract}

\section{Introduction}

The discovery of \emph{novel} small-molecule drugs is of paramount significance in medicine. Drug-like molecules reside in a search space with a size estimated to be up to $10^{60}$~\citep{wayne1996}. The size of this space far exceeds the capacity of current \textit{in vitro} assays and \textit{in silico} virtual screening methods. This leads to a 'needle-in-a-haystack' problem, where the challenge lies in finding drug molecules with the desired optimal properties that reside in an exponentially small subspace. Even state-of-the-art high-throughput experimental methods such as DNA-encoded small molecule libraries~\citep{liu_del_2011, mayr_hts_2009, llyod_hts_2020, reiher_del_2021} and in silico ultra high-throughput molecular docking~\citep{gorgulla2020virtualflow, shoichet2019ultra, shoichet2020docking,Lyu2023AlphaFold2ST,Gorgulla2023VirtualFlow2} still do not go beyond a search space of $10^{10}$ molecules. Moreover, all these techniques explore a limited, and often biased subset of the chemical space. Indeed, past research has shown that the chemical diversity and size of chemical libraries are directly correlated with hit-rate and potency of identified hits. Development of enabling technologies that allow expanding the searchable chemical space is therefore of crucial importance in drug development. ~\citep{Lyu2023ModelingTE}

\begin{figure}
    \centering
    \begin{subfigure}[t]{0.95\textwidth}
    \includegraphics[width=0.95\columnwidth]{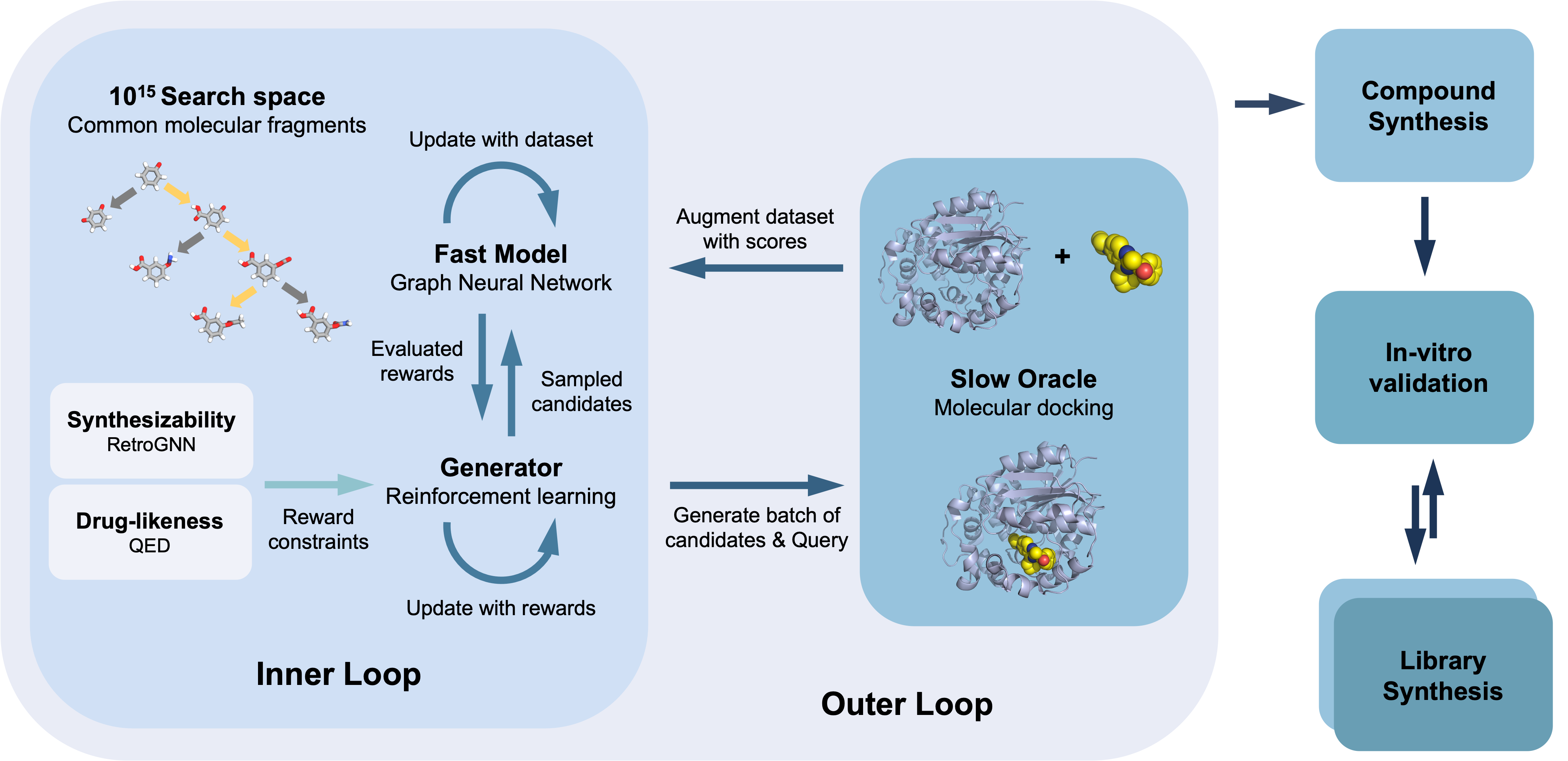}
    \end{subfigure}
    \caption{Schematics of \lz illustrating the overall approach. The approach consists of learning a fast surrogate model which is used to guide a generative policy to design \emph{de novo} molecules with constraints on synthesizability and drug-likeness. Batches of candidates generated with the policy are evaluated using the molecular docking oracle. This whole outer loop is executed for a few rounds enriching the library. We then select candidates for in vitro synthesis and validation.}
    \label{fig:schematics}
\end{figure}

Machine learning approaches have the potential to accelerate the exploration of molecular space as they can use few samples to generalize to unseen chemicals by learning the underlying latent structure. Supervised ML models can approximate biochemical properties, such as antimicrobial activities or molecular docking scores, and when coupled with virtual screening in a known chemical space, this leads to a significant speed-up~\citep{cherkasov2020deepdocking, regina2020antibiotics, Abe2023DeepLD}. However, the speed gain offered by these methods, while significant ($\sim10^3-10^5\times$ faster than traditional methods), is insufficient in face of the exponentially large search space of size $10^{60}$. Beyond virtual screening by enumeration, local optimization methods are able to find good local extrema in structure-activity space through a series of small improvements, without visiting every molecule. For example, a synthon-based virtual optimization strategy has been shown to search a space equivalent to 11 billion molecules while only evaluating 100 million molecules~\citep{sadybekov2022synthon}. Beyond local optimization, active learning strategies that employ iterative sampling and improvement are particularly sample efficient in virtual screening~\citep{bhat2019activelearning,jerome2021activelearning} and have been shown to filter libraries of up to 138 million compounds~\citep{coley2021activelearnning}. To explore truly \textit{de novo} generation, reinforcement learning algorithms can learn generative policies to search efficiently in combinatorially large spaces, making them a natural fit. Indeed, several \textit{de-novo} generative approaches have yielded compounds validated \textit{in vitro}, for example in identifying ligands with $IC_{50} < 100 nM$~\citep{aspuruguzik2019gentrl, Korshunova2022GenerativeAR, Merk2018TuningAI, yang2022generative}. However, current \textit{de novo} generation of small molecule inhibitors are faced with the trilemma of generating molecules too similar to the training set (i.e. insufficient exploration), ensuring compound synthesizability, and defining a cheap and reasonable objective function~\citep{Gao2020TheSO}.

In this paper, we introduce \lz, a generative active learning approach that \textit{learns to search} in the combinatorially large space of small molecules with constraints on synthesizability. \lz consists of a generative policy trained with reinforcement learning, an approximate Bayesian surrogate model which models a computationally expensive docking simulation, a synthesizability model~\citep{liu2022retrognn} and a drug-likeliness model~\citep{bickerton2012qed}, and an acquisition function to guide the search. To demonstrate the effectiveness of the approach, we apply \lz to the generation of small molecule protein inhibitors for a clinically relevant drug target, the enzyme soluble Epoxide Hydrolase 2 (sEH)~\citep{imig2009seh, mccloskey2020del}. With only $\sim10^4$ docking simulations, \lz produces synthesizable, drug-like molecules with docking scores that would otherwise require the virtual screening of a hundred billion ($10^{11}$) molecules. We selected a molecule with a scaffold that has not been observed in known sEH inhibitors and synthesized a set of analogs around this scaffold. Experimental validation showed that the best performing molecule inhibited sEH with submicromolar potency. 

\section{Learning to search in molecular space}

There are three key challenges in \textit{in silico} molecular design: a) the estimation of important biochemical properties, such as binding affinity through docking simulations, are computationally expensive, b) the space of small molecules is combinatorially large, and c) ensuring the synthesizability and drug-likeliness of the designed molecules is difficult. The various components of \lz, illustrated in Figure~\ref{fig:schematics} and discussed in further detail in Appendix~\ref{app:algo_details}, aim to address each of these challenges. We note that while we study the design of sEH binders, our method is applicable to any property where a scoring function is available.

To search for sEH binders, we utilized docking scores from Autodock Vina~\citep{arthur2010autodock}. Validation against a set of sEH inhibitors from the ChEMBL database with experimental binding affinity showed that this score is moderately inversely correlated with experimental IC\textsubscript{50} (Spearman's rank correlation $r_s\sim0.4$)~\citep{mendez2019chembl}. This docking simulation, however, takes roughly 5-6 minutes to evaluate a single proposed molecule on CPU. Using this score directly for searching in the large space of molecules is intractable. 

To address this, we leverage a surrogate model to approximate the biochemical property of interest. Specifically, we use a pre-trained $E(n)$ invariant graph neural network~\citep{satorras2021n} to model the docking score. The graph neural network is pre-trained on a dataset of 200,000 docked molecules from Zinc to improve the out-of-distribution prediction performance. In a held-out random validation set, the model's normalized MAE is $\sim0.3$; for validation sets split by scaffold or docking scores, the normalized MAE increases to $\sim0.6-0.7$. This out-of-distribution performance is critical to enable exploration of novel chemical spaces. The surrogate model is able to approximate the docking at a fraction of the computational cost. 

The other key component of \lz is the generative policy which is trained to maximize the property of interest. The policy operates in the space of chemical building blocks, comprised of the 131 most common fragments extracted from the PDB ligand database, which can be combined with discrete actions that connect or disconnect two of these building blocks~\citep{jin2020hierarchical, bengio2021flow, liu2022retrognn}. This subset of small molecule space contains up to $10^{15}$ molecules and is biased towards drug-like molecules. The policy is trained using proximal policy optimization ~\citep[PPO;][]{schulman2017ppo} algorithm with entropy regularization and count-based rewards to improve exploration~\citep{Ahmed2018UnderstandingTI, tang_exploration_2017}. Constructing the molecule through a sequence of steps adding fragments enables the policy to generalize to unseen molecules which share the same fragment. 

Simply maximizing the docking score with the generative policy without additional constraints always resulted in highly unfeasible molecules. To ensure the designed molecules are synthesizable and drug-like we incorporate soft constraints via the reward function of the generative policy. The reward for the generative policy is a combination of the score with a drug-likeness score QED~\citep{bickerton2012qed} and a synthetic accessibility score estimated by a RetroGNN~\citep{liu2022retrognn} which is trained using retrosynthetic analysis. This reward design guides the policy to generate molecules which are likely to be synthesizable while being strong binders.

The inner loop of \lz comprises of training the generative policy to maximize the reward defined by the surrogate model along with QED and RetroGNN. Since the surrogate model is imperfect, a batch of the top-scoring candidates generated by the policy is then evaluated by the expensive molecular docking simulation, forming the outer loop. The resultant docking scores are augmented to the dataset, and the inner loop restarts by improving the surrogate model on the augmented dataset. Over iterations of the outer loop, \lz enriches the library of generated molecules to higher docking scores, and the best performing candidates are identified for \textit{in vitro} validation.

\section{Results}

\begin{figure}[htbp]
    \centering
    \begin{subfigure}{0.49\textwidth}
    \includegraphics[width=1\linewidth]{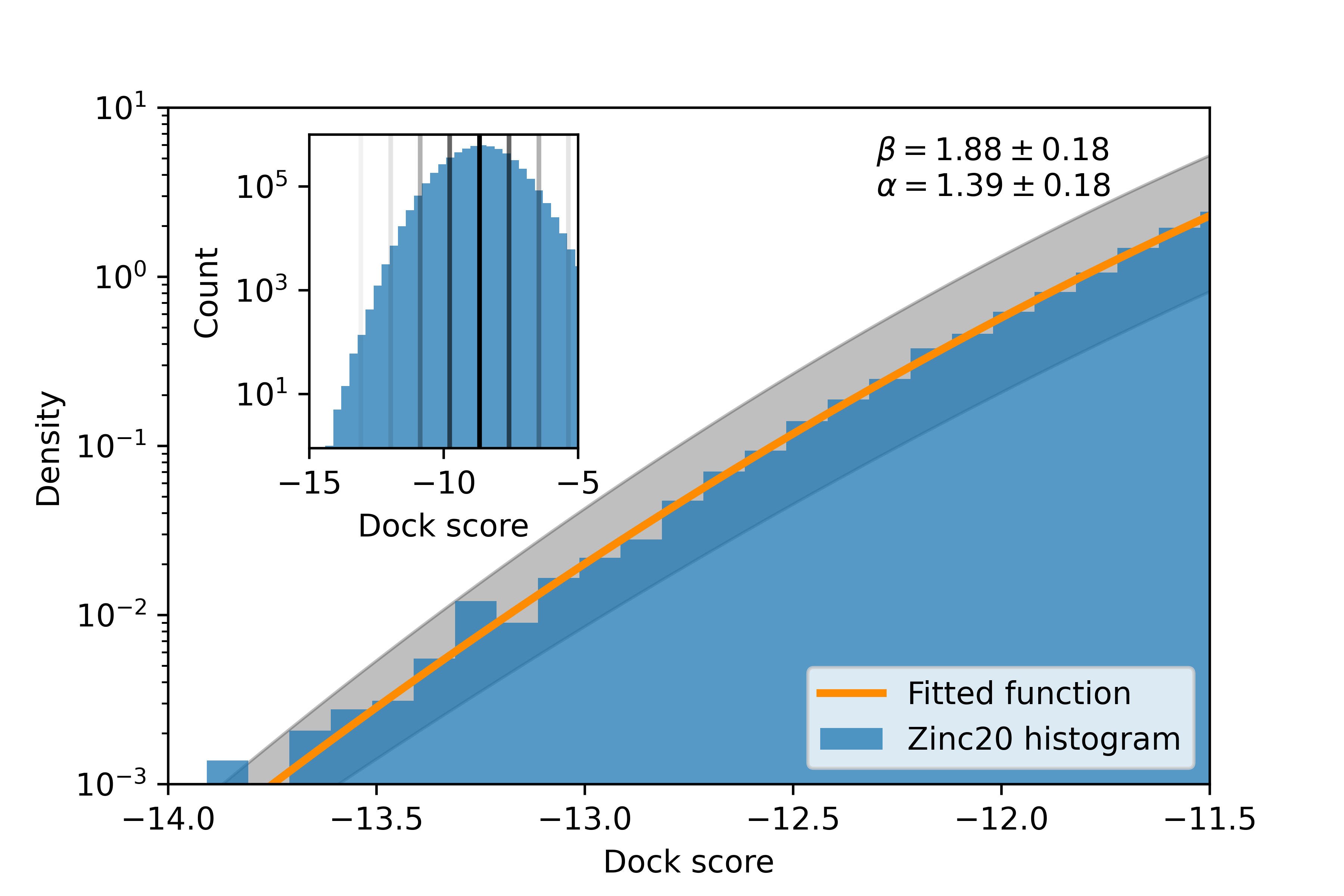}
    \caption{}        
    \label{fig:zinc_distribution}
    \end{subfigure}
    \begin{subfigure}{0.49\textwidth}
    \includegraphics[width=0.9\linewidth]{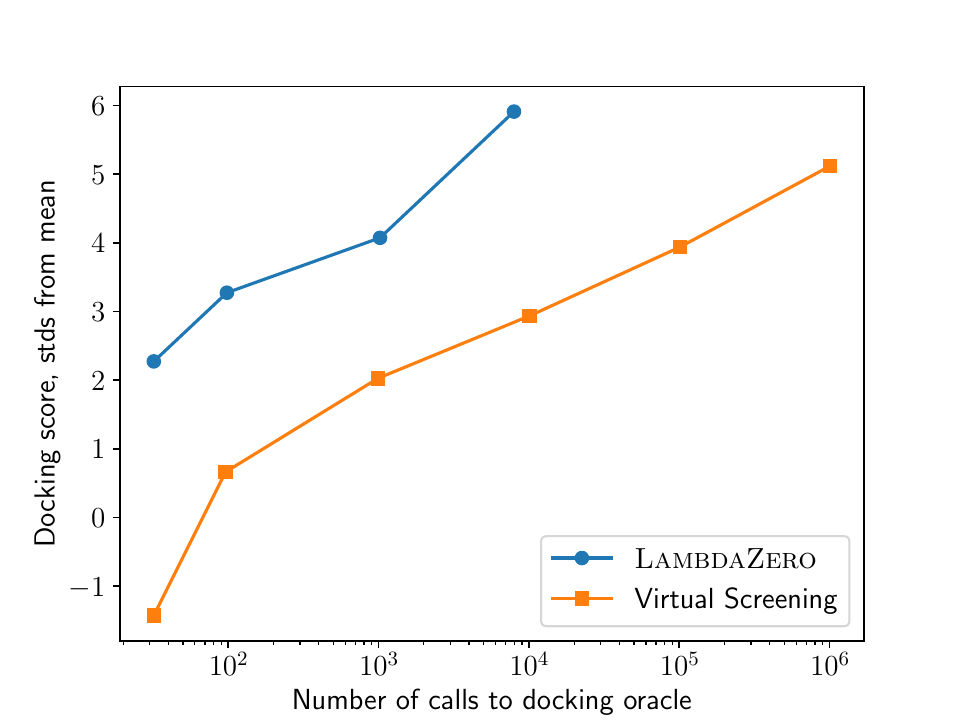}
    \caption{}
    \label{fig:lambdazerosearch}
    \end{subfigure}
    \caption{\lz searches exponentially faster than virtual screening. (a) The tail of the distribution ($x >\mu+2.5\sigma$) of 5.8 million dock scores of drug-like molecules from Zinc20 with a generalized Gaussian distribution fit and its 95\% confidence interval. The inset shows the remaining distribution with mean and $\pm1, 2, 3 \sigma$. (b) The number of calls to oracles against the highest reached normalized docking scores for LambdaZero and virtual screening in Zinc dataset.} 
    \label{fig:insilico_results}
\end{figure}

\begin{figure}[htbp]
    \centering
      \begin{minipage}[c][][t]{.55\textwidth}
    \begin{subfigure}{\textwidth}
      \centering
      \includegraphics[width=0.9\textwidth]{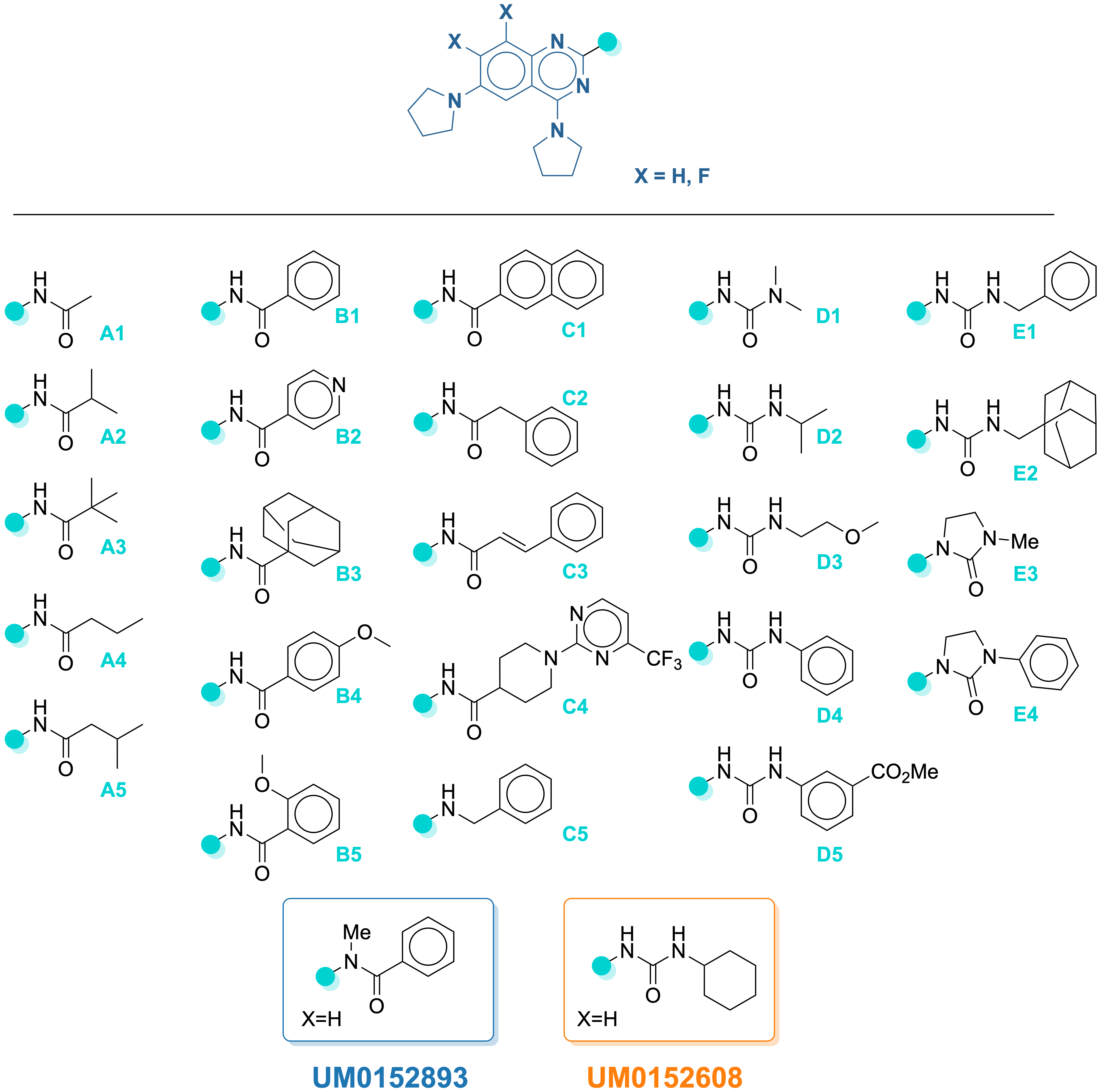}
      \caption{}
      \label{fig:synthesized_mols}
    \end{subfigure}
  \end{minipage}%
  \begin{minipage}[c][][t]{.4\textwidth}
    \begin{subfigure}[t]{\textwidth}
      \centering
      \includegraphics[width=0.7\textwidth]{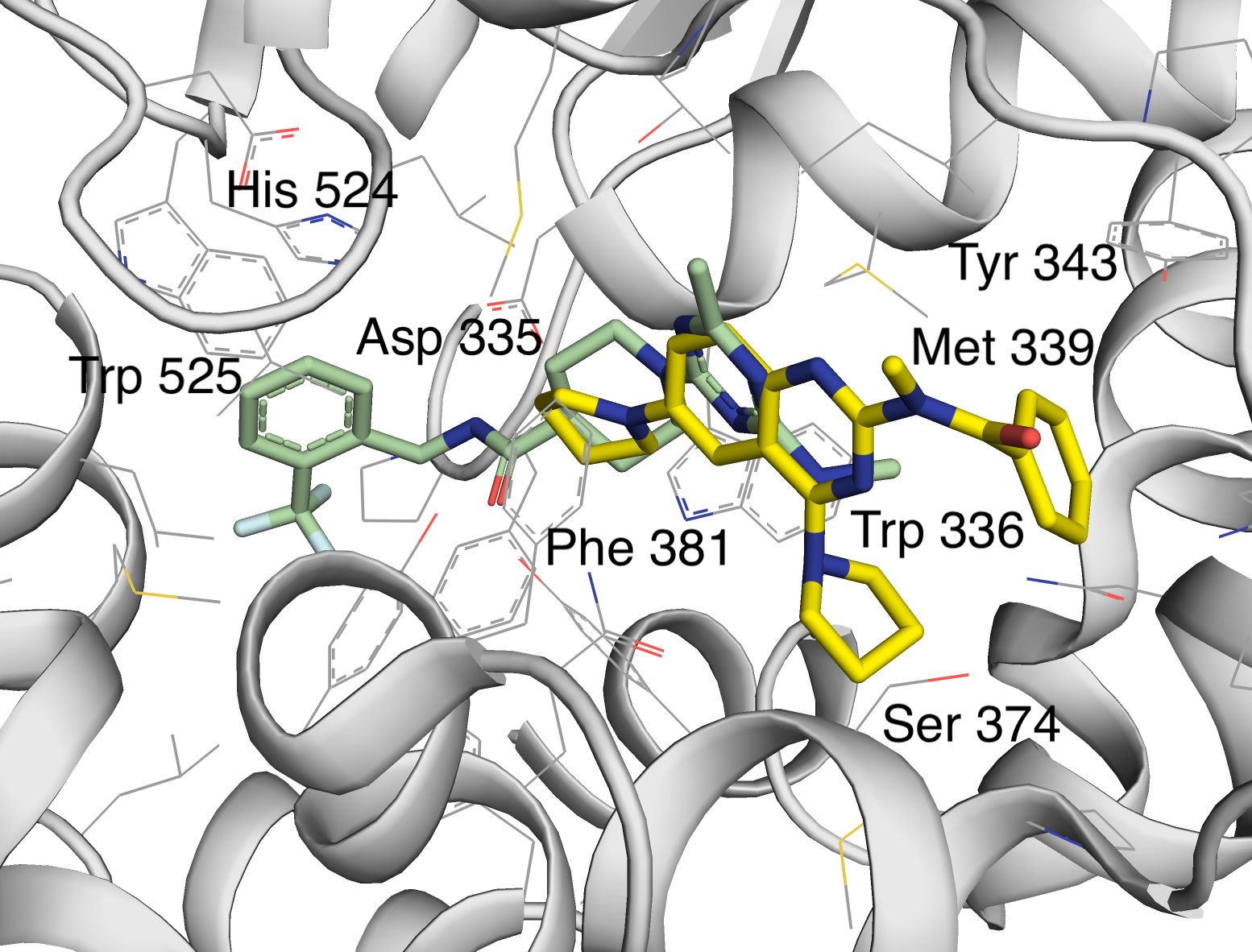}
      \caption{}
      \label{fig:dockingpose}
    \end{subfigure}
    \vfill
    \begin{subfigure}[t]{\textwidth}
      \centering
      \includegraphics[width=0.7\textwidth]{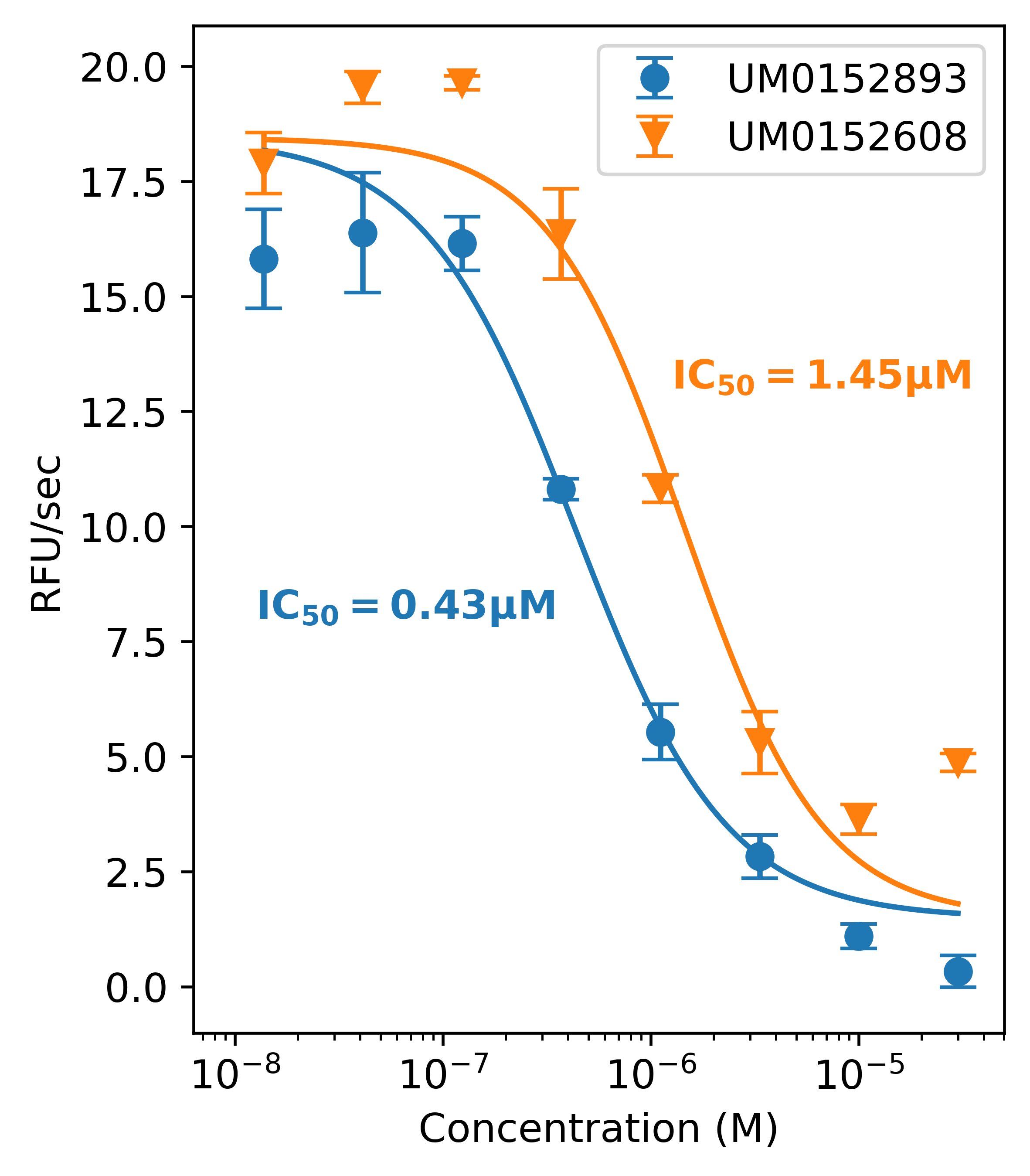}
      \caption{}
      \label{fig:ic50}
    \end{subfigure}
  \end{minipage}
    \caption{\lz designs leads to synthesizable sEH protein inhibitors. (a) Synthesized molecular library based on scaffold discovered by \lz, and the highlighted strongest inhibitors. (b) Docking pose of UM0152608 (yellow), compared to the native sEH ligand in PDB 4jnc (green). Selected sEH amino acid residues in contact with the ligands have been labeled. (c) Concentration-response curves of top two compounds and calculated IC\textsubscript{50} values. Data are plotted as mean ± standard deviation of three replicates.}
    \label{fig:experimental_validation}
\end{figure}

\xhdr{In silico results.} To assess efficiency of searching the molecular space with \lz, we first performed docking on sEH with 5.8 million molecules from the Zinc20 drug-like molecule dataset~\citep{irwin2020zinc}. The distribution of the docking scores (Figure~\ref{fig:zinc_distribution}) is mostly Gaussian ($\mu=-8.67$, $\sigma=1.10$), where the tail ($x>\mu+2.5\sigma$) is fitted to a generalized Gaussian distribution with $\beta=1.88\pm0.18$ and $\alpha=1.38\pm0.18$. The cumulative distribution function (CDF) of this function can be used to evaluate the progress of search algorithms by computing the number of expected molecules to virtually screen to reach the same score. 

Figure~\ref{fig:lambdazerosearch} illustrates the performance of \lz relative to a naive virtual screening baseline from drug-like molecules in Zinc 20 database. As the number of calls to oracle grows in the log scale, the top docking scores increase linearly in terms of standard deviations from the mean of the Zinc20 dataset. We observe that the active learning process and better policies are crucial in reducing oracle calls: the curve corresponding to \lz has a steeper gradient, and reaches a higher score in a small fraction of docking oracle calls. A linear approximation to this curve indicates that to reach a given score, the number of queries $n$ is exponentially fewer with better search algorithms, i.e. $n_1 = n_2^{m_2/m_1}$, where $m$ denotes the slope of the curve. \lz can reach up to a z-score of 6.75 (-16.1) with only $\sim10^4$ docking queries. Approximately, this would have required docking $\sim10^{11}$ molecules from Zinc20. 

The molecules generated by \lz generally have QED$>0.7$ and RetroGNNScore$<4.5$, maintaining drug-likeliness and synthesizability. The molecules generated are chemically distinct from the Zinc20 dataset as well as from known sEH inhibitors (Figure~\ref{fig:tanimoto}). 

\xhdr{Experimental validation of molecular docking. }To obtain empirical validation of the use of the AutoDock VINA docking algorithm as an oracle in \lz design of sEH inhibitors, we purchased 25 high scoring molecules selected from the virtual screening of 5.8 million commercially available drug-like molecules from Zinc-15 (vide supra). We measured their sEH inhibitory activity in an \textit{in vitro} enzyme inhibition assay alongside a series of known sEH inhibitors. Eight out of the twenty-five Zinc-15 compounds displayed sEH inhibiting activity at 10 $\mu$M, and all were also new sEH inhibitors. The most potent compound from the Zinc dataset had an IC\textsubscript{50} of 0.4 $\mu$M, with the rest having an IC\textsubscript{50} ranging from 0.7-40 $\mu$M  (3 compounds with an IC\textsubscript{50} $<1\mu$M). In short, this demonstrates that molecular docking can be employed as an (imperfect) oracle in the design process of sEH inhibitors. 

\xhdr{Experimental validation of \lz designs. }\textit{De novo} sEH inhibitor designs generated by \lz were subsequently evaluated by medicinal chemists on several metrics including frequency of appearance in top-100 molecules, drug-likeness, synthetic feasibility, availability of starting materials, scaffold novelty, and potential for an analogue library. We selected a scaffold core based on  N-(4,6-di(pyrrolidin-1-yl)quinazolin-2-yl) amide. Compared to the native ligand in PDB 4jnc~\citep{thaji2013sehbinder}, this scaffold is computationally predicted to occupy a different position in the binding pocket (Figure~\ref{fig:dockingpose}). We synthesized a small library of  35 analogue molecules in milligram scales (see Appendix~\ref{app:synth_details} for synthesis details). We note that an additional 49 proposed analogs suffered from poor solubility and purification issues and were not synthesized. Out of all synthesized molecules, 24  of the 35 displayed inhibition activity against sEH in an in-vitro enzyme inhibition assay (see Appendix~\ref{app:exp_details}. IC\textsubscript{50} values were calculated from the dose response curves and ranged from approximately 90 $\mu$M to 0.4  $\mu$M (Figure~\ref{fig:ic50}, Table~\ref{tab:hydrocompounds},~\ref{tab:monofluorocompounds},~\ref{tab:difluorocompounds}); specifically, 2 molecules were identified to have an IC\textsubscript{50} of $\sim 1 \mu$M (UM0152608, $pIC_{50}=5.838$ with 95\% CI of $5.951$ to $5.713$), and $\sim 0.4 \mu$M (UM0152893, $pIC_{50} = 6.365$ with 95\% CI of $6.488$ to $6.248$). The most potent variant, N-(4,6-di(pyrrolidin-1-yl)quinazolin-2-yl)-N-methylbenzamide (UM0152893), had a relatively small amide substituent, and larger substituents at this position were also relatively potent (in the micromolar range). The core scaffold is not observed from the virtual screening in Zinc20, and to the best of our knowledge, the core scaffold is novel amongst known sEH inhibitors.

\section{Discussion}
We present \lz, a generative active learning approach composed of a fast surrogate model, a generative policy with synthesizability and drug-likeness constraints, and an expensive computational oracle. \lz can efficiently design novel, synthesizable small-molecule protein binders in a fragment-based molecule search space. \lz demonstrates exponential gains over existing methods allowing the equivalent virtual screening of $10^{11}$ molecules. 

We apply \lz to design binders for the enzyme sEH. A chosen scaffold was experimentally synthesized, the analogue library of which contained 24/35 inhibitors with potencies down into the $\sim$ submicromolar regime. We note that various components of \lz can be replaced by recent alternatives such as graph transformers~\citep{masters2023gps++} for the surrogate model and GFlowNets~\citep{bengio2021flow} for training the generative policy. We also note that to approximate protein binding, we used molecular docking; while docking correlates with experimental results (both via retrospective and prospective study), it likely is the bottleneck of \lz to finding highly effective protein binders. We expect our work, in conjunction with higher-fidelity oracles, search spaces of more easily synthesizable molecules, and other pharmacological considerations (e.g. ADMET), to enable the fast discovery of high affinity small molecule protein binders. 

\section*{Acknowledgements}
We thank Jason Hartford, John Bradshaw, Paul Maragakis, Joanna Chen, and Arnaud Bergeron for helpful discussions and feedback. 

We acknowledge funding from Génome Québec, CQDM Fonds d’accélération des collaborations en santé (FACS) / Acuité Québec, Canadian Institutes of Health Research (CIHR), Intel and Anyscale. 

The research was enabled in part by computational resources provided by the Digital Research Alliance of Canada (\url{https://alliancecan.ca}), Mila (\url{https://mila.quebec}), Intel and NVIDIA.

\section*{Disclosure}

\noindent{\small Optimization Notice: Software and workloads used in performance tests may have been optimized for performance only on Intel microprocessors. Performance tests, such as SYSmark and MobileMark, are measured using specific computer systems, components, software, operations and functions. Any change to any of those factors may cause the results to vary. You should consult other information and performance tests to assist you in fully evaluating your contemplated purchases, including the performance of that product when combined with other products. For more information go to \url{http://www.intel.com/performance}. Intel, Xeon, and Intel Xeon Phi are trademarks of Intel Corporation in the U.S. and/or other countries.}

\bibliography{iclr2024_conference}
\bibliographystyle{iclr2024_conference}

\clearpage
\appendix
\section{Algorithmic Details}
\label{app:algo_details}
\subsection{Additional Details}
Algorithm~\ref{algo:overall} summarizes the overall \lz pipeline. The approach consists of an outer loop of optimization interacting with the molecular docking program and an inner loop which consists of fitting a surrogate model on the collected data and training the generative policy using the surrogate model along with soft constraints to generate candidates to be evaluated with molecular docking in the outer loop. 

\begin{algorithm}[h]
\SetAlgoLined
\kwInput{\\
$\calO$: Computational oracle to evaluate candidates $x$ and return labels $Y$\;
$D_0=\{(x_i, y_i)\}$: Initial dataset with prior evaluations of the oracle, $y_i=\mathcal{O}(x_i)$\;
$\calM$: Surrogate model estimating $E[Y|x]$ 
$\pi$: Generative policy trainable using a reward function $R$ used to generate potential candidates $x$\;
$b$: Size of candidate batch to be generated\; 
$N$: Number of outer loop iterations\;
$K$: Number of top-scoring candidates to be synthesized and evaluated\;}
\textbf{Procedure:}\\
\For{$i=1$ to $N$}{
$\bullet$ Fit surrogate model $\calM$ on dataset $D_{i-1}$\;
$\bullet$ Train generative policy $\pi$ using reward function $R$\;
$\bullet$ Construct a batch of candidates sampled from the generative policy, $B = \{x_1, \dots, x_b\}$ where $x_i \sim \pi$\;
$\bullet$ Evaluate batch $B$ with $\calO$: \mbox{$\hat{D_i}=\{(x_1, \calO(x_1)), \dots, (x_b, \calO(x_b))\}$}\;
$\bullet$ Update dataset $D_i=\hat{D_i} \cup D_{i-1}$\;
}
\KwResult{$TopK(D_N)$ elements $(x,y) \in D_n$ with highest values of $y$}

\caption{\lz: Overall Approach}
\label{algo:overall}
\end{algorithm}

\subsection{Surrogate Model}
\paragraph{Architecture}
We use a $E(n)$ invariant graph neural network~\citep{satorras2021n} as the base architecture for the surrogate model. We initially tested a variant of MPNN~\citep{gilmer2017mpnn} but observed significantly better performance with the EGNN. We use a hidden dimension of 128, with 6 layers. 

\paragraph{Training}
We first pretrain the model on a subset of the Zinc20 dataset. To construct this dataset, we first run a docking simulation on all the molecules in the dataset. Then we bin the molecules in dataset according to their docking score. We then sample uniformly from the bins to get the final subset of interest. This set has $200,000$ molecules. We train the EGNN on this set with a standard regression objective for 10 epochs using the Adam optimizer and learning rate of $0.0025$ with a batch size of $64$.

\subsection{Generative Policy}
\paragraph{Environment} 
For the generative policy we use the fragment-based molecule generation environment from \citet{bengio2021flow}. We start with a library of 131 molecular fragments extracted from the PDB ligand database following~\citet{jin2020hierarchical}. This selection of fragments from the PDB ligand library serves as a strong prior on the policy to explore a subspace of the overall molecule space biased towards synthesizability. Each molecule in the space covered by these fragments can be generated through a sequence of actions (trajectory) combining these fragments. We limit the generation to molecules which can be constructed in 7 steps, and we cap the maximum number of non-hydrogen atoms at 50. This process defines a MDP, where the states are the (partially constructed) molecule and actions at each state correspond to the possible blocks that can be added to the existing molecule or stopping the generation at that step. This is a deterministic MDP, with variable number of actions available at each step, with rewards only available at the terminal state. Additionally during generation to ensure diversity in the generated molecules, we take 3 random steps at the beginning of each trajectory to encourage diversity in the generation.

\paragraph{Reward Design} Design of the reward function used for training the policy is critical to ensure the synthesizability of the designed molecules. Let $s(x)$ denote the docking score estimated by the surrogate model, and $qed(x)$ and $synth(x)$ denote the QED score for drug likeness and RetroGNN score for synthesizability respectively. To compute these scores we first define cutoffs $\min qed$ and $\min synth$ on the QED and synthesizability score respectively. Scores below these cutoffs are mapped to 0, and above the cutoffs are normalized to $[0,1]$ where the max is defined a priori. The reward for the policy is then defined as follows:
\begin{equation}
    R(x) = s(x) * qed(x) * synth(x)
\end{equation}

This formulation corresponds to ``soft'' constraints on the QED and synthesizability scores. It encourages the generation of molecules which simultaneously have a high score from the surrogate model, high drug-likeness measured by the QED score and high synthesizability as measured by the RetroGNN score. 
\paragraph{Policy Architecture}
For the generative policy we use a message passing neural network~\citep[MPNN;][]{gilmer2017mpnn} as the base architecture.The MPNN has a hidden dimension of 128, and 6 layers of message passing. We use a softmax policy, parameterized by logits from the base MPNN. The MPNN produces logits for all possible actions at each step, but we mask the logits to only allow the sampling of valid actions at each step. 

\paragraph{Training}
We train the policy using proximal policy optimization~\citep[PPO;][]{schulman2017ppo}. We choose PPO to train the policy primarily owing to its flexibility and generally strong performance in various RL benchmarks. In particular, we use the efficient multi-GPU PPO implementation from Ray RLLib~\citep{liang2018rllib}\footnote{\url{https://docs.ray.io/en/latest/rllib/rllib-algorithms.html\#ppo}}. Additionally, in order to improve the exploration, we use entropy regularization~\citep{Ahmed2018UnderstandingTI} and count-based exploration~\citep{tang_exploration_2017}. Within each outer loop step, we train the policy for $10,000$ steps, with a learning rate of $0.0001$ with $4$ PPO epochs per step, using $16$ concurrent environment instantiations for data collection, and an entropy coefficient of $0.05$.

\subsection{Implementation Details}

Given the scale of the search space, efficient scalability of the implementation is critical. In this section we discuss several optimizations we have made to ensure efficient execution of LambdaZero on large clusters.

\paragraph{Resource} Due to the scale of the project, we utilized multiple different compute environments. The initial experiments were conducted with 4 NVIDIA V100 GPUs with 40 CPU cores in 2 Intel\textsuperscript{\textregistered} Gold 6148 Skylake CPUs and 128GB RAM on a single node. In further experiments we used 16 dual-CPU nodes each consisting of $2$ Intel\textsuperscript{\textregistered} Xeon\textsuperscript{\textregistered}~ Platinum 8480+ (Sapphire Rapids) CPUs with $56$ cores each. 

\subsubsection{Key CPU optimizations}
\paragraph{Surrogate model} The surrogate model is based on an Equivariant Graph Neural Network (EGNN) formulation. EGNN training time consists of two stages: edge feature computation including update and aggregation (AGG) and a subsequent node-feature update. The first stage consists of three components: initial edge-feature computation as distance between source and destination nodes, updating edge features via a Multi Layer Perceptron (Edge MLP) and finally, their Aggregation onto destination node features. In the second stage, we apply a Node MLP to perform node-feature update. We use Deep Graph Library (DGL)~\citep{dgl} for AGG operations since it provides hardware efficient implementations of various GNN algorithms ~\citep{md2021distgnn}. DGL’s AGG implementation uses multi-threading and SIMD and is built on top of the LIBXSMM library ~\citep{libxsmm}. The LIBXSMM library provides highly hardware efficient SIMD based primitives for many matrix operations. It provides more instruction reduction than manually written SIMD intrinsics based code by using JITing to generate optimal assembly code with SIMD instructions. We also rely on LIBXSMM to optimize update operations; we create fused, tiled and multi-threaded operations of the Linear operator ($h^l = h^{l-1} . W^{l-1} + b^l$) and the following activation function (Sigmoid Linear Unit or SiLU). These changes result in a $2.24\times$ speedup in training the surrogate model. 

\paragraph{Generative Policy} This model consists of a sequence of feedforward layers (among other operations) each consisting of a linear transform followed by a LeakyReLU activation. We fuse these two operators and apply tiling to ensure optimized implementation of these operations. We observe that this step of the overall training algorithm scales with number of compute threads and adjust the threading mechanism accordingly. These optimizations accelerate the training of the generative policy on a single CPU by $4.62\times$.

\paragraph{Molecular Docking} The molecular docking oracle is a critical bottleneck in the entire computational pipeline. We parallelize it across multiple dual-CPUs nodes. Molecular docking is embarrassingly parallel, in the sense that each molecule can be docked independently from one another. We distribute the molecules equally among the dual-CPU nodes. In \lz, we use AutoDock-Vina for molecular docking. Since in AutoDock-Vina, default value of “exhaustiveness” is set to $8$, it does not scale beyond $8$ cores for one molecule. Therefore, we use $8$ cores for each molecule. On each dual-CPU node with $112$ cores over two Sapphire Rapids CPUs, we run $\frac{112}{8} = 14$ instances of AutoDock-Vina  at a time to dock $14$ molecules in parallel. Since different molecules may require different amount of time for docking, we use dynamic scheduling across the $14$ instances running on a dual-CPU node.

With all optimizations put together, we accelerate \lz by $13.12 \times$ from $20.17$ hours to nearly $1.54$ hours per step. This enables us to reduce the total execution time of $25$ steps of the outer loop from $21$ days to just $1.6$ days.

\section{Experimental Details}
\label{app:exp_details}
\subsection{Molecular Docking and Scoring}

The molecular docking protocol for Soluble Epoxide Hydrolase (sEH) was conducted using AutoDock Vina, which involved preprocessing a set of 5.8 million diverse, drug-like molecules from the ZINC20 database. RDKit was employed to convert SMILES representations to 3D-coordinates and exported to .mol file format, ensuring accurate structural data for docking. Following this conversion, the 3D structure of the C-terminal  sEH domain of Bifunctional epoxide hydrolase 2 in complex with a carboxamide inhibitor was used (obtained from the PDB 4jnc) was used as docking receptor.  Structure preparation of the receptor was performed according to standard practices with the binding site for docking defined $\sim$9\r{A} around the coordinates of the naive carboxamide inhibitor. AutoDock Vina then calculated docking scores to predict the binding affinities between the sEH and each molecule, with high-scoring conformations identified as potential inhibitors.

\begin{figure}[htbp]
    \centering
      \captionsetup[subfigure]{labelformat=empty}
    \begin{subfigure}[t]{\textwidth}
          \centering
          \includegraphics[width=0.6\textwidth]{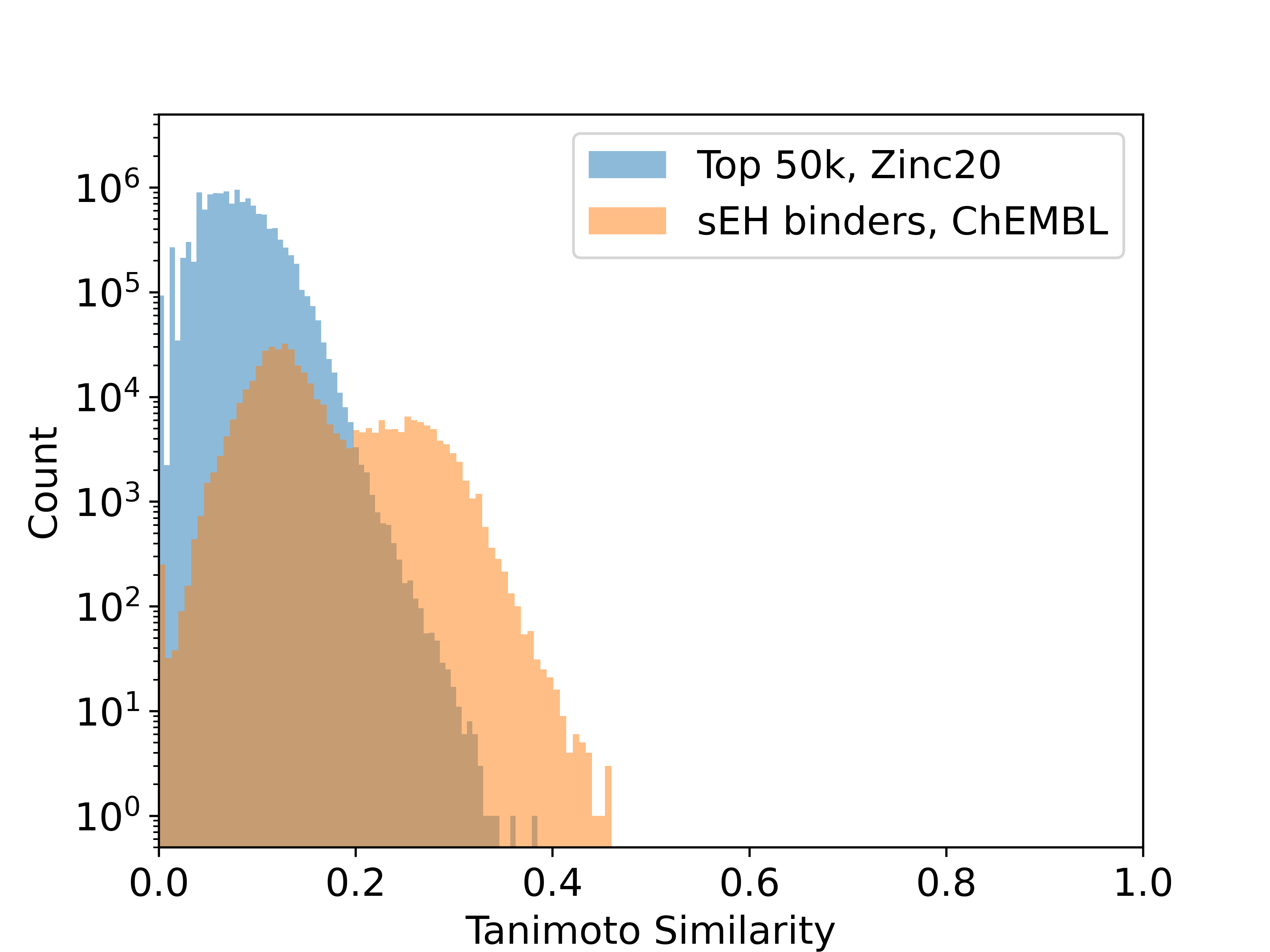}
        \end{subfigure}
    \caption{The distribution of pairwise Tanimoto molecular similarity between \lz generated molecules and known sEH inhibitors from ChEMBL and 50,000 molecules with highest docking score from virtual screening in Zinc20.}
    \label{fig:tanimoto}
\end{figure}

\subsection{In vitro sEH assay}

The in vitro soluble epoxide hydrolase (sEH) assay was adapted from the method described by Litovchick et al~\citep{litovchick2015encoded}. Epoxy-fluor-7 substrate (cat \#10008610) and sEH (cat\# 10011669) were obtained from Cayman Chemical. The assay was performed in an assay buffer consisting of 25 mM Tris pH 7 and 0.01 mg/mL BSA, with a final enzyme concentration of 3 nM and substrate concentration of 2 $\mu$M in low-volume Greiner 384 well plates, with a total assay volume of 20 $\mu$L. Compounds were dissolved in DMSO and dispensed by either an acoustic dispenser (Echo) or manually, at concentrations of 30 $\mu$M and 10 $\mu$M for compounds, or 1 $\mu$M for t-AUCB control. For dose response curves, a serial 3-fold dilution series was constructed ranging from 30 $\mu$M to 0.2 nM. The compounds and enzyme were combined in half the assay volume, and after a 20-minute incubation at room temperature, the EF-7 substrate was added. The rate of substrate conversion was measured for 40 minutes by continuously monitoring fluorescence (excitation at 330 nm and emission at 465 nm wavelengths) at 30°C using a multimode plate reader (Tecan M1000). The rates were calculated for the linear portion of the curve between 10 seconds and 500 seconds and were fit to a 4-parameter sigmoid curve to obtain the IC50.

\section{Synthetic details}
\label{app:synth_details}

\subsection{General Synthetic Methods and Materials:}
Commercially available chemicals and solvents were used as received without further purification. Reactions were performed under a nitrogen atmosphere and all glassware was dried and purged with N\textsubscript{2} before use. Organic solutions were concentrated under reduced pressure on a rotary evaporator using a water bath. Reactions were monitored with TLC, analytical LCMS, and/or \textsuperscript{1}H NMR. Preparative reverse-phase high-pressure liquid chromatography was carried out on a preparative LCMS Agilent 1260 Infinity II series equipped with an Agilent 6120 Quadrupole LC/MS mass spectrometer and C18 columns using methanol/water gradients containing 0.1\% acetic acid. Collected fractions were concentrated as described above or on a Genevac HT-4X. LCMS analysis was performed on an Agilent 1260 Infinity II series. All new compounds gave satisfactory \textsuperscript{1}H NMR and LCMS results and all final compounds had an LCMS purity of >95\% in the case of single compounds and >80\% in the cases of compounds obtained from parallel synthesis (unless otherwise indicated). LCMS analyses were performed using the following conditions: Analytical LCMS Method A. Column: Kinetex-C18 2.6$\mu$m, 3.0×30 mm. Mobile phase: A = MeOH:H\textsubscript{2}O:HCO\textsubscript{2}H (5:95:0.05\%), B = MeOH:H\textsubscript{2}O:HCO\textsubscript{2}H (95:5:0.05\%). \textsuperscript{1}H NMR were obtained on a Varian 400 MHz or a Bruker Ascend 400 MHz using the residual signal of deuterated NMR solvent as internal reference.

\subsection{New compound synthesis and characterization}

\begin{figure}[htbp]
    \centering
      \begin{minipage}[c][][t]{.9\textwidth}
      \captionsetup[subfigure]{labelformat=empty}
    \begin{subfigure}{\textwidth}
      \centering
      \includegraphics[width=0.9\textwidth]{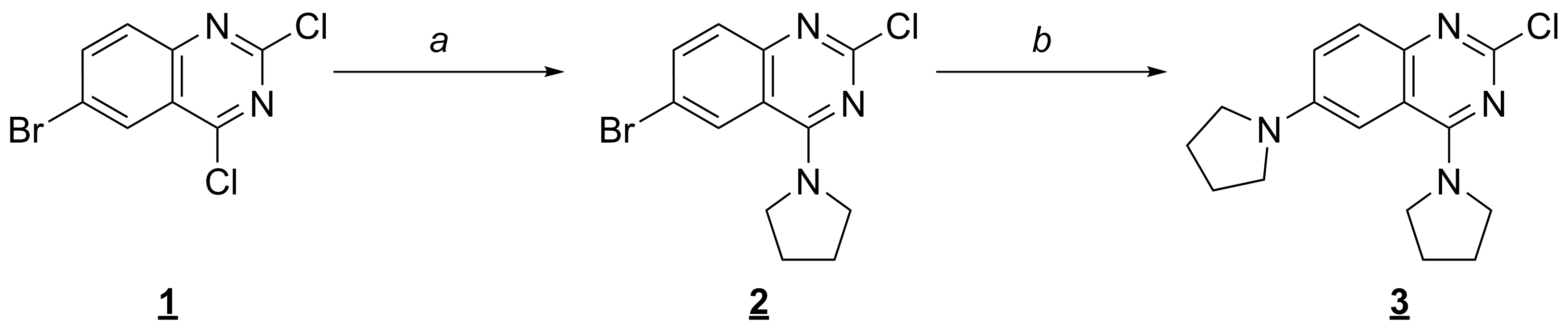}
      \caption{\textit{Reagents}: \textit{a}. pyrrolidine, DIPEA, NMP, room temperature; \textit{b}. pyrrolidine, XanPhos, NaOBu-t, Pd\textsubscript{2}(dba)\textsubscript{3}}
    \end{subfigure}
  \end{minipage}%
    \caption{Synthesis of quinazoline-based scaffold \textbf{\underline{(3)}}.}
    \label{fig:synthesis_1}
\end{figure}

\xhdr{6-Bromo-2-chloro-4-(pyrrolidin-1-yl)quinazoline \underline{(2)}}: To a solution of  6-bromo-2,4-dichloro-quinazoline (1) (2.00 g, 7.20 mmol) in NMP (20 mL) was added pyrrolidine (0.517 g, 7.27 mmol), followed by  N,N-diisopropylethylamine (1.40 g, 1.89 mL, 10.8 mmol), and the mixture was stirred at room temperature for 30 min. The reaction mixture was then diluted with water and saturated aqueous NaHCO\textsubscript{3} was added to give a precipitate. The precipitate was filtered and the filter-cake was washed with water and then dried in vacuo overnight to afford 6-bromo-2-chloro-4-(pyrrolidin-1-yl)quinazoline (2) (2.11 g, 94\%) as a beige solid. LCMS: 1.281 min; (M+H)\textsuperscript{+} = 311.8. \textsuperscript{1}H NMR (400 MHz, DMSO-d\textsubscript{6}) $\delta$ (ppm) 8.37 (d, \textit{J}= 2.0 Hz, 1H), 7.92 (m, 1H), 7.57 (d, \textit{J}= 9.0 Hz, 1H), 3.88 (m, 4H), 1.97 (m, 4H).

\xhdr{2-Chloro-4,6-di(pyrrolidin-1-yl)quinazoline \underline{(3)}}: A 20 ml microwave vial was charged with 6-bromo-2-chloro-4-(pyrrolidin-1-yl)quinazoline (2) (1.20 g, 3.84 mmol), pyrrolidine (0.287 g, 4.03 mmol), (9,9-dimethyl-9H-xanthene-4,5-diyl)bis(diphenylphosphine) (0.133 g, 0.23 mmol) and sodium t-butoxide (0.553 g, 5.76 mmol). The vial was purged with N\textsubscript{2}, toluene (12 mL) was added and the resulting suspension was purged with a stream of N\textsubscript{2} bubbles for 5 min. To this mixture was added tris(dibenzylideneacetone)dipalladium (0) (0.070 g, 0.077 mmol) and the mixture was again purged with N\textsubscript{2} for 5 min. The reaction vial was sealed and the mixture was heated at 105°C (oil-bath temperature) for 15 h. The cooled mixture was partitioned with DCM-water, the organic phase was separated and the aqueous phase was extracted with DCM (x3). The combined organic extract was washed (water, brine), dried (Na\textsubscript{2}SO\textsubscript{4}) and evaporated to give a solid. The crude solid was taken up in DCM and pre-adsorbed on silica gel to prepare a sample cartridge. Flash chromatography (40 g silica gel column) eluting with 0-60\% ethyl acetate-DCM afforded 2-chloro-4,6-di(pyrrolidin-1-yl)quinazoline (3) (0.64 g, 55\%) as a light yellow solid. LCMS: 1.420 min; (M+H)\textsuperscript{+} = 303.0.\textsuperscript{1}H NMR (400 MHz, CDCl\textsubscript{3}) $\delta$ (ppm) 7.65 (d, \textit{J}= 9.4 Hz, 1H), 7.09 (m, 2H), 3.98 (m, 4H), 3.37 (m, 4H), 2.06 (m, 8H).

\begin{figure}[htbp]
    \centering
      \begin{minipage}[c][][t]{.95\textwidth}
      \captionsetup[subfigure]{labelformat=empty}
    \begin{subfigure}{\textwidth}
      \centering
      \includegraphics[width=0.9\textwidth]{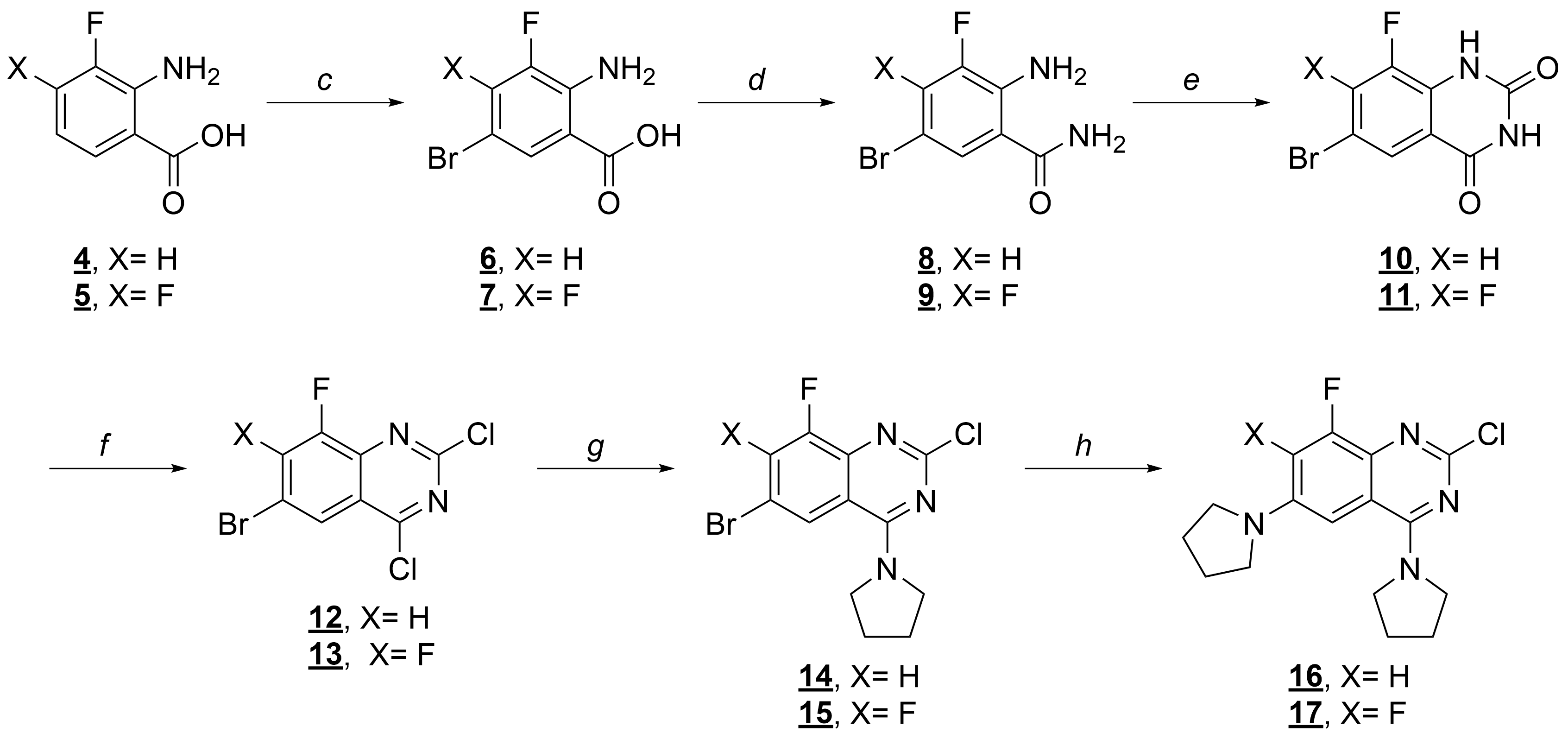}
      \caption{\textit{Reagents}: \textit{c}. NBS, DCM; \textit{d}. NH\textsubscript{4}OH, HATU, DIPEA, DMF; \textit{e}. triphosgene, DIPEA, DCM, 40°C; \textit{f}. POCl\textsubscript{3}, DIPEA, 110°C; \textit{g}. pyrrolidine, DIPEA, NMP; \textit{h}. pyrrolidine, Pd\textsubscript{2}(dba)\textsubscript{3}, XantPhos, NaOBu-t, toluene, 105°C.}
    \end{subfigure}
  \end{minipage}%
    \caption{Synthesis of fluoro-quinazoline-based scaffolds (\textbf{\underline{16}} and \textbf{\underline{17}}).}
    \label{fig:synthesis_2}
\end{figure}

\xhdr{2-Amino-5-bromo-3-fluorobenzoic acid \underline{(6)}}: To a solution of 2-amino-3-fluorobenzoic acid (4) (5.00 g, 32.2 mmol) in DCM (50 mL) was added N-bromosuccinimde (5.74 g, 32.2 mmol) and the mixture was stirred at room temperature overnight. After 18 h, the resulting mixture was filtered, and the filter-cake was washed with dichloromethane (3 x 75 mL). The resulting solid was dried under reduced pressure to obtain 2-amino-5-bromo-3-fluoro-benzoic acid (6) (6.22 g, 82\%) as a solid. LCMS: 1.201 min; (M-H)\textsuperscript{-} = 231.8. \textsuperscript{1}H NMR (400 MHz, DMSO-d\textsubscript{6}) $\delta$ (ppm) 7.51 (dd, \textit{J}= 10.6, 2.4 Hz, 1H) 7.62 (dd, \textit{J}= 2.4, 1.6 Hz, 1H).

\xhdr{2-Amino-5-bromo-3,4-difluorobenzoic acid \underline{(7)}}: The product was obtained from compound (5), using the method described for the synthesis of compound (6). LCMS: 1.230 min; (M-H)\textsuperscript{-} = 249.8 .\textsuperscript{1}H NMR (400 MHz, DMSO-d\textsubscript{6}) $\delta$ (ppm) 13.27 (br s, 1H), 7.76 (dd, \textit{J}= 7.4, 2.4 Hz, 1H).

\xhdr{2-Amino-5-bromo-3-fluorobenzamide \underline{(8)}}: To a solution of 2-amino-5-bromo-3-fluorobenzoic acid (6) (13.42 g, 57.3 mmol) and HATU (30.5 g, 80.0 mmol) in DMF (224 mL) was added triethylamine (15.99 mL, 115 mmol), followed by concentrated ammonium hydroxide (23.92 mL, 172 mmol). After the mixture was stirred for 1 h at room temperature, then water (1 L) was added and the mixture was extracted with DCM (3 x 150 mL). The combined organic extracts were evaporated under reduced pressure and the residue was dried in vacuo overnight. This afforded the product (8) (10.20 g, 76\%) as a solid which was used as such in the next step. LCMS: 1.099 min; (M-H\textsubscript{2}O)+ = 215.8. \textsuperscript{1}H NMR (400 MHz, DMSO-d\textsubscript{6}) $\delta$ (ppm) 7.96 (br s, 1H), 7.62 (m, 1H), 7.42 (m, 2H), 6.64 (s, 2H).

\xhdr{2-Amino-5-bromo-3,4-difluorobenzamide \underline{(9)}}: The product was obtained from compound (7), using the method described for the synthesis of compound (8). LCMS: 1.125 min; (M+H)\textsuperscript{+} = 250.9 .\textsuperscript{1}H NMR (400 MHz, DMSO-d\textsubscript{6}) $\delta$ (ppm) 7.98 (br s, 1H), 7.76 (dd, \textit{J}= 7.0, 2.4 Hz, 1H), 7.40 (br s, 1H), 6.98 (s, 2H).

\xhdr{6-Bromo-8-fluoroquinazoline-2,4(1H,3H)-dione \underline{(10)}}: To a mixture of 2-amino-5-bromo-3-fluorobenzamide (8) (2.00 g, 8.58 mmol) in DCM (80 mL) was added triphosgene (2.73 g, 9.18 mmol), followed by DIPEA (1.49 mL, 8.58 mmol). The reaction vessel was sealed and the mixture was heated at 40°C (oil-bath temperature) overnight. After 18 h, the cooled mixture was filtered to give a white powder. This material was taken up in DCM-MeOH and adsorbed onto silica gel. The volatiles were removed under reduced pressure and the solid was dry-packed into a sample cartridge. Flash chromatography (40 g silica gel column), eluting with 0-100\% ethyl acetate-hexanes and then 0-10\% MeOH-DCM afforded the desired product (10) (1.63 g, 74\%) as a white powder. LCMS: 1.094 min; (M-H)\textsuperscript{-} = 256.8.\textsuperscript{1}H NMR (400 MHz, DMSO-d\textsubscript{6}) $\delta$ (ppm) 7.89 (dd, \textit{J}= 10.2, 2.4 Hz, 1H), 7.78 (dd, \textit{J}= 2.4, 1.2 Hz, 1H), 7.62 (s, 1H), 7.00 (s, 1H).

\xhdr{6-Bromo-7,8-difluoroquinazoline-2,4(1H,3H)-dione \underline{(11)}}: The product was obtained from compound (9), using the method described for the synthesis of compound (10). LCMS: 1.246 min; (M-H)\textsuperscript{-} = 274.8. \textsuperscript{1}H NMR (400 MHz, DMSO-d\textsubscript{6}) $\delta$ (ppm) 11.70 (s, 1H), 11.60 (s, 1H), 7.92 (m, 1H).

\xhdr{6-Bromo-2,4-dichloro-8-fluoroquinazoline \underline{(12)}}: To a solution of 6-bromo-8-fluoro-1H-quinazoline-2,4-dione (10) (0.200 g, 0.77 mmol) in POCl\textsubscript{3} (1.1 mL) was added DIPEA (0.200 mL, 1.16 mmol). The reaction vessel was sealed and the mixture was heated at 110°C overnight. After 18 h, the cooled mixture was diluted with DCM and then the volatiles were removed under reduced pressure. The crude mixture was taken up in DCM (30 mL) and water (15 mL) was carefully added. The organic phase was separated and the aqueous phase was re-extracted with DCM (2 x 30 mL). The combined organic extract was washed (saturated aqueous NaHCO\textsubscript{3}, brine), dried (Na\textsubscript{2}SO\textsubscript{4}) and evaporated to give the crude product as a brown solid. This solid was taken up in DCM and dry-packed on silica gel. The volatiles were removed under reduced pressure and the solid was transferred to a sample cartridge. Flash chromatography (4 g silica gel column) eluting with 0-10\% ethyl acetate-hexanes afforded the title compound (12) (0.159 g, 70\%) as an off-white solid. LCMS: 1.277 min; (M+H)\textsuperscript{+} = 294.8. \textsuperscript{1}H NMR (400 MHz, CDCl\textsubscript{3}) $\delta$ (ppm) 8.24 (t, \textit{J}= 1.6 Hz, 1H), 7.83 (dd, \textit{J}= 8.6, 2.0 Hz, 1H).

\xhdr{6-Bromo-2,4-dichloro-7,8-difluoroquinazoline \underline{(13)}}: The product was obtained from compound (11), using the method described for the synthesis of compound (12). LCMS: 1.298 min; (M-F)- = 294.9. \textsuperscript{1}H NMR (400 MHz, CDCl\textsubscript{3}) $\delta$ (ppm) 8.37 (dd, \textit{J}= 6.3, 2.4 Hz, 1H).

\xhdr{6-Bromo-2-chloro-8-fluoro-4-(pyrrolidin-1-yl)quinazoline \underline{(14)}}: The product was obtained from compound (12), using the method described for the synthesis of compound (2). LCMS: 1.393 min; (M+H)\textsuperscript{+} = 329.8.\textsuperscript{1}H NMR (400 MHz, CDCl\textsubscript{3}) $\delta$ (ppm) 8.07 (t, \textit{J}= 1.8 Hz, 1H), 7.56 (dd, \textit{J}= 9.0, 2.0 Hz, 1H) 3.95 (m, 4H), 2.10 (m, 4H).

\xhdr{6-Bromo-2-chloro-7,8-difluoro-4-(pyrrolidin-1-yl)quinazoline \underline{(15)}}: The product was obtained from compound (13), using the method described for the synthesis of compound (2). LCMS: 1.297 min; (M+H)\textsuperscript{+} = 347.8. \textsuperscript{1}H NMR (400 MHz, DMSO-d\textsubscript{6}) $\delta$ (ppm) 8.32 (dd, \textit{J}= 6.7, 2.4 Hz, 1H), 3.86 (m, 4H), 1.96 (m, 4H).

\xhdr{2-Chloro-8-fluoro-4,6-di(pyrrolidin-1-yl)quinazoline \underline{(16)}}: The product was obtained from compound (14), using the method described for the synthesis of compound (3). LCMS: 1.470 min; (M+H)\textsuperscript{+} = 320.9. \textsuperscript{1}H NMR (400 MHz, DMSO-d\textsubscript{6}) $\delta$ (ppm) 7.08 (dd, \textit{J}= 13.3, 2.3 Hz, 1H), 6.91 (d, \textit{J}= 1.9 Hz, 1H), 3.89 (br s, 4H), 3.34 (br s, 4H), 1.97 (m, 8H).

\xhdr{2-Chloro-7,8-difluoro-4,6-di(pyrrolidin-1-yl)quinazoline \underline{(17)}}: The product was obtained from compound (15), using the method described for the synthesis of compound (3). LCMS: 1.489 min; (M+H)\textsuperscript{+} = 339.0.\textsuperscript{1}H NMR (400 MHz, CDCl\textsubscript{3}) $\delta$ (ppm) 6.99 (dd, \textit{J}= 8.2, 1.6 Hz, 1H), 3.95 (m, 4H), 3.48 (m, 4H), 2.04 (m, 8H).

\begin{figure}[htbp]
    \centering
      \begin{minipage}[c][][t]{.9\textwidth}
      \captionsetup[subfigure]{labelformat=empty}
    \begin{subfigure}{\textwidth}
      \centering
      \includegraphics[width=0.9\textwidth]{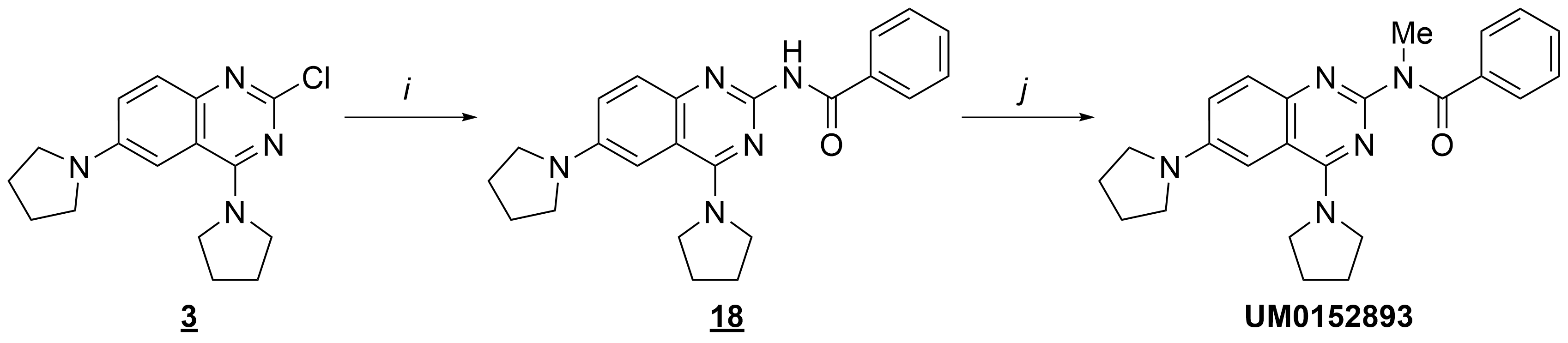}
      \caption{\textit{Reagents}: \textit{i}. benzamide, Cs\textsubscript{2}CO\textsubscript{3}, XPhos, Pd(OAc)\textsubscript{2}, DMF, 100°C; \textit{j}. iodomethane, Cs\textsubscript{2}CO\textsubscript{3}, DMF, 45°C.}
    \end{subfigure}
  \end{minipage}%
    \caption{Synthesis of amide-based analog \textbf{UM0152893}}
    \label{fig:synthesis_3}
\end{figure}

\xhdr{General Method A. N-(4,6-Di(pyrrolidin-1-yl)quinazolin-2-yl)benzamide \underline{(18)}}: A reaction vial was charged with 2-chloro,4,6-di(pyrrolidin-1-yl)quinazoline (3) (0.030 g, 0.099 mmol), benzamide (0.018 g, 0.15 mmol), cesium carbonate (0.058 g, 0.18 mmol), 2-dicyclohexylphosphino-2',4',6'-tri-isopropyl-1,1'-biphenyl (0.014 g, 0.030 mmol) and palladium (II) acetate (0.0022 g, 0.0099 mmol). The vial was sealed and then purged with N\textsubscript{2}, then DMF (1 mL) was added and the mixture was purged with a stream of N\textsubscript{2} bubbles for 1 min. The resulting mixture was then heated at 100°C (oil-bath temperature) for 10 h. The cooled mixture was filtered, the filtrate was diluted to 2 mL with DMSO and the solution was acidified with a few drops of AcOH. This mixture was purified by Prep-LCMS (Reverse-phase Kinetex 5µm C18 column 21.2 x 100 mm; elution with MeOH-water-0.1\% AcOH. Gradient: Isocratic 30\% for 1.5 minutes, then gradient to 100\% MeOH over 10 minutes.). The product-containing fractions were combined and evaporated and the residue was lyophilized from MeCN-water to afford N-(4,6-di(pyrrolidin-1-yl)quinazolin-2-yl)benzamide (18) (0.014 g, 36\%) as a light yellow solid. LCMS: 1.349 min; (M+H)\textsuperscript{+} = 388.0. \textsuperscript{1}H NMR (400 MHz, DMSO-d\textsubscript{6}): $\delta$ (ppm) 12.00 (s, 1H), 8.07 (d, \textit{J}= 9.3 Hz, 1H), 8.06 (s, 1H), 8.00 (d, \textit{J}= 9.3 Hz, 1H), 7.72 (m, 1H), 7.61 (t, \textit{J}= 7.0 Hz, 2H), 7.31 (dd, \textit{J}= 9.3, 2.8 Hz, 1H), 7.18 (d, \textit{J}= 2.3 Hz, 1H), 4.33 (m, 2H), 3.91 (m, 2H), 3.36 (m, 4H), 2.08 (m, 8H).

\xhdr{N-(4,6-Di(pyrrolidin-1-yl)quinazolin-2-yl)-N-methylbenzamide \underline{(UM0152893)}}: To a mixture of N-(4,6-di(pyrrolidin-1-yl)quinazolin-2-yl)benzamide (18) (8.5 mg, 0.022 mmol) and cesium carbonate (7.1 mg, 0.022 mmol) in DMF (0.6 mL) was added iodomethane (6.2 mg, 0.044 mmol). The reaction vessel was sealed under N\textsubscript{2} and the mixture was heated at 45°C (oil-bath temperature) for 2 h. A second portion of iodomethane (6.2 mg, 0.044 mmol) was then added and heating was continued at the same temperature for 18 h. Another portion of iodomethane (6.2 mg, 0.044 mmol) was then added and heating was continued at 55°C for 3 h. The cooled mixture was then partitioned with DCM-water, the organic phase was separated and the aqueous phase was re-extracted with DCM (x2). The combined organic extract was washed (brine), dried (Na\textsubscript{2}SO\textsubscript{4}) and concentrated. The concentrate is taken up in MeOH-DMSO-water (total volume 2 mL) and purified by Prep-LCMS (Reverse-phase Kinetex 5µm C18 column 21.2 x 100 mm; eluted with MeOH- water-0.1\% formic acid, gradient to 100\% MeOH). The product-containing fractions were combined and evaporated and the residue was lyophilized from MeCN-water to afford N-(4,6-di(pyrrolidin-1-yl)quinazolin-2-yl)-N-methylbenzamide (UM0152893) (1.0 mg, 10\%) as a yellow solid. LCMS: 1.353 min; (M+H)\textsuperscript{+} = 402.2. \textsuperscript{1}H NMR (400 MHz, CDCl\textsubscript{3}): $\delta$ (ppm) 7.55 (d, \textit{J}= 8.5 Hz, 1H), 7.43 (d, \textit{J}= 7.1 Hz, 1H), 7.02 – 7.26 (m, 4H), 6.97 (s, 1H), 3.70 (s, 3H), 3.43 (br s, 4H), 3.34 (br s, 4H), 2.05 (br s, 4H), 1.80 (br s, 4H).

\begin{figure}[htbp]
    \centering
      \begin{minipage}[c][][t]{.7\textwidth}
      \captionsetup[subfigure]{labelformat=empty}
    \begin{subfigure}{\textwidth}
      \centering
      \includegraphics[width=0.9\textwidth]{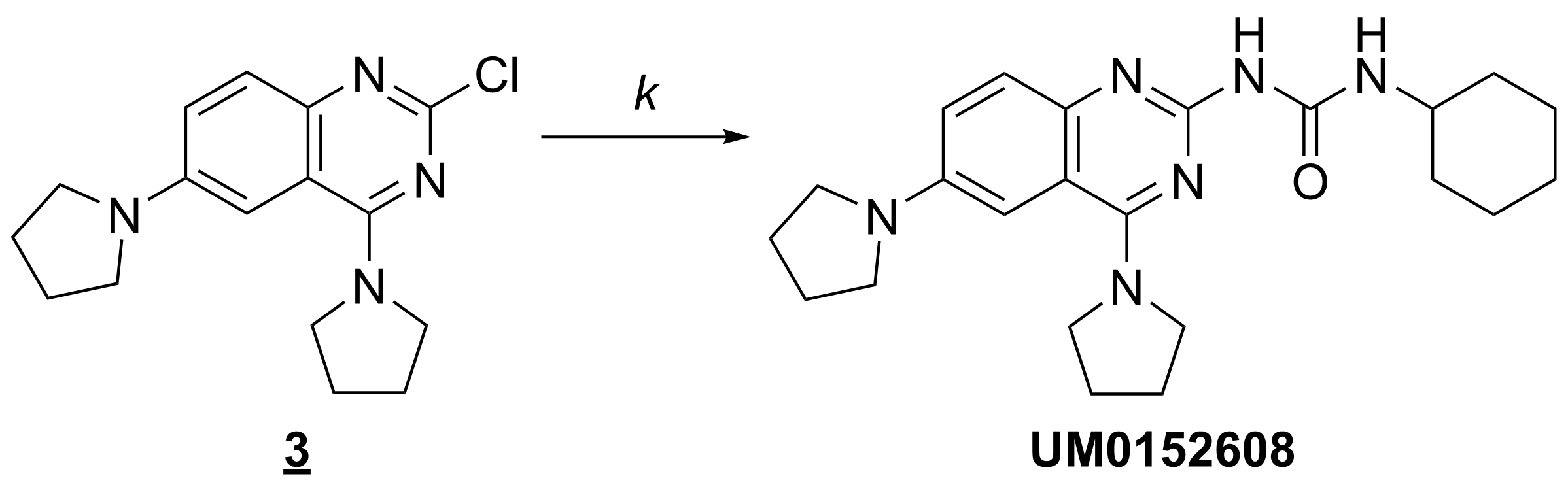}
      \caption{\textit{Reagents}: \textit{k}. cyclohexylurea, Cs\textsubscript{2}CO\textsubscript{3}, XPhos, Pd(OAc)\textsubscript{2}, DMF, 100°C.}
    \end{subfigure}
  \end{minipage}%
    \caption{Synthesis of urea-based analog \textbf{UM0152608}}
    \label{fig:synthesis_4}
\end{figure}

\xhdr{General Method B. 1-Cyclohexyl-3-(4,6-di(pyrrolidin-1-yl)quinazolin-2-yl)urea \underline{(UM0152608)}}: A reaction vessel was charged with 2-chloro,4,6-di(pyrrolidin-1-yl)quinazoline (3) (0.040 g, 0.13 mmol), cyclohexylurea (0.028 g, 0.20 mmol), cesium carbonate (0.077 g, 0.24 mmol), 2-dicyclohexylphosphino-2',4',6'-tri-isopropyl-1,1'-biphenyl (0.019 g, 0.040 mmol) and palladium (II) acetate (0.003 g, 0.013 mmol). The vial was sealed and then purged with N2, then DMF (2 mL) was added and the mixture was purged with a stream of N\textsubscript{2} bubbles for 1 min. The resulting mixture was then heated at 85°C (oil-bath temperature) for 15 h. The cooled mixture was partitioned with DCM-water, the organic phase was separated and the aqueous phase was re-extracted with DCM (x2). The combined organic extract was washed (water, brine), dried (Na\textsubscript{2}SO\textsubscript{4}) and concentrated. The concentrate was diluted with toluene and evaporated (done twice) and the residue was taken up in DCM and pre-adsorbed on silica gel. The volatiles were removed under reduced pressure and the solid was transferred to a sample cartridge. Flash chromatography (4 g silica gel column), eluting with 0-30\% (DCM-MeOH, 8:2)-DCM, afforded the impure product. Flash chromatography of this impure solid was repeated using the same conditions to afford material that was still contaminated. The impure product was taken up in MeOH-DMSO-water and purified by Prep-LCMS (Reverse-phase Kinetex 5µm C18 column 21.2 x 100 mm; elution with MeOH-water-0.1\% AcOH, gradient to 100\% MeOH). The product-containing fractions were combined and evaporated and the residue was lyophilized from MeCN-water to give 1-cyclohexyl-3-(4,6-di(pyrrolidin-1-yl)quinazolin-2-yl)urea (UM0152608) (0.005 g, 9\%) as a solid. LCMS: 1.488 min; (M+H)\textsuperscript{+} = 409.2. \textsuperscript{1}H NMR (400 MHz, CDCl\textsubscript{3}): $\delta$ (ppm) 9.65 (br s, 1H), 7.44 (d, \textit{J}= 8.9 Hz, 1H), 6.95 – 7.15 (m, 2H), 3.93 (m, 4H), 3.80 (br s, 1H), 3.34 (m, 4H), 2.06 (m, 8H), 1.77 (dt, \textit{J}= 13.4, 3.9 Hz, 2H), 1.62 (dd, \textit{J}= 8.4, 4.1 Hz, 1H), 1.24 – 1.49 (m, 8H).

\begin{figure}[htbp]
    \centering
      \begin{minipage}[c][][t]{.7\textwidth}
      \captionsetup[subfigure]{labelformat=empty}
    \begin{subfigure}{\textwidth}
      \centering
      \includegraphics[width=0.9\textwidth]{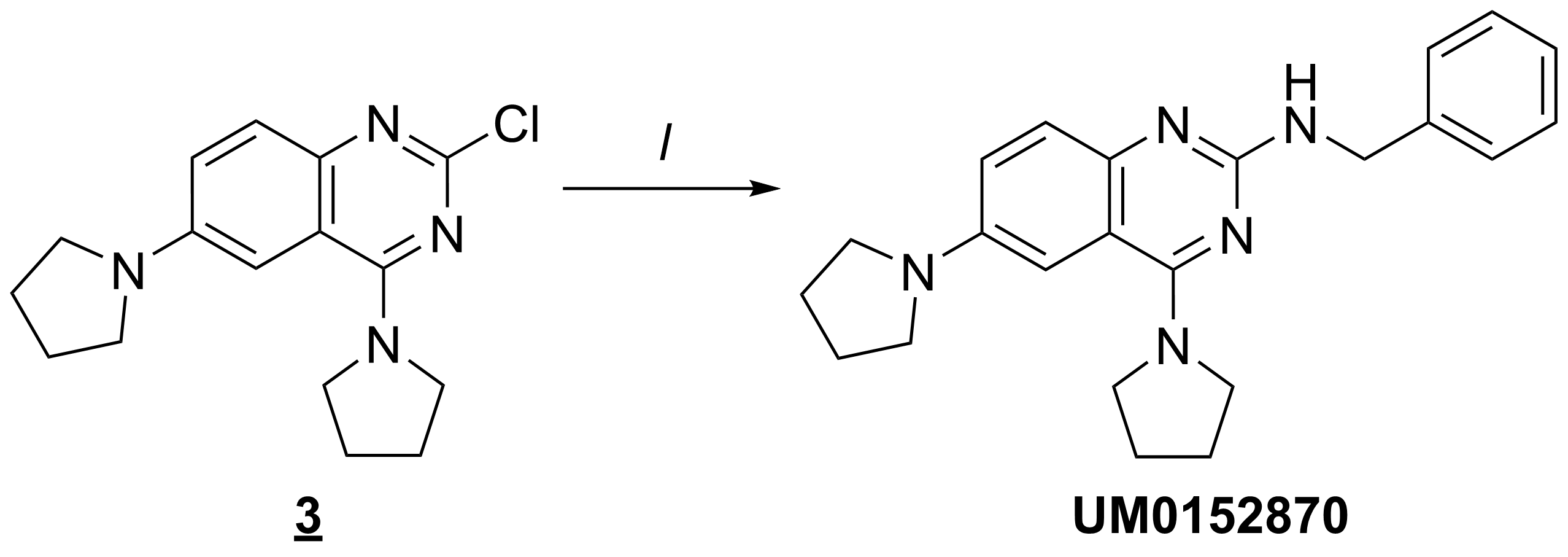}
      \caption{\textit{Reagents}: \textit{l}. benzylamine, pentanol, 140°C.}
    \end{subfigure}
  \end{minipage}%
    \caption{Synthesis of amine-based analog \textbf{UM0152870}. }
    \label{fig:synthesis_5}
\end{figure}

\xhdr{General Method C. N-Benzyl-4,6-dipyrrolidin-1-yl-quinazolin-2-amine \underline{(UM0152870)}}: To a solution of 2-chloro-4,6-dipyrrolidin-1-yl-quinazoline (3) (10 mg, 0.033 mmol) in 1-pentanol (0.33 mL) was added benzylamine (3.5 mg, 0.033 mmol), the reaction vessel was capped and the mixture was stirred at 140°C (oil-bath temperature) for 15 h. The cooled mixture was diluted with a MeOH-DMSO-water mixture and purified by Prep-LCMS (Reverse-phase Kinetex 5µm C18 21.2 x 100 mm column; elution with MeOH-water-0.1\% HCO\textsubscript{2}H). The product-containing fractions were evaporated and the residue was lyophilized from MeCN-water to give N-benzyl-4,6-dipyrrolidin-1-yl-quinazolin-2-amine (1 mg, 8\%) as a beige solid. LCMS: 1.498 min; (M+H)\textsuperscript{+} = 374.0. \textsuperscript{1}H NMR (400 MHz, CDCl\textsubscript{3}): $\delta$ (ppm) 7.48 (d, \textit{J}= 9.0 Hz, 1H), 7.40 (d, \textit{J}= 7.8 Hz, 2H), 7.31 (t, \textit{J}= 7.1 Hz, 2H), 7.24 (d, \textit{J}= 7.0 Hz, 1H), 7.01 (d, \textit{J}= 9.0 Hz, 1H), 6.98 (s, 1H), 4.67 (s, 2H), 3.92 (m, 4H), 3.31 (m, 4H), 2.04 (m, 8H).


\begin{table}[htbp]
\centering
\caption{Compound Characterization Data of \textit{Hydrogenated} Quinazoline Compounds (4,6-di(pyrrolidin-1-yl)quinazoline). Reference to Figure~\ref{fig:synthesized_mols} for the corresponding compound, where $X = H$} 
\begin{tabularx}{\textwidth}{l|X|l}
\toprule
\textbf{Compound} & \textbf{Analytical Data} & \textbf{IC\textsubscript{50} ($\mu$M)} \\
\midrule
UM0152893 & \textsuperscript{1}H NMR (400 MHz, CDCl\textsubscript{3}): $\delta$ (ppm) 1.80 (br. s., 4 H) 2.05 (br. s., 4 H) 3.34 (br. s., 4 H) 3.43 (br. s., 4 H) 3.70 (s, 3 H) 6.97 (s, 1 H) 7.02 - 7.26 (m, 4 H) 7.43 (d, \textit{J}=7.13 Hz, 2 H) 7.55 (d, \textit{J}=8.51 Hz, 1 H). LCMS \textit{m/z}: [M+H]\textsuperscript{+} Calcd for C\textsubscript{24}H\textsubscript{28}N\textsubscript{5}O 402.2; Found 402.2. & 0.43 \\
\midrule
UM0152608 & \textsuperscript{1}H NMR (400 MHz, CDCl\textsubscript{3}): $\delta$ (ppm) 1.20 - 1.50 (m, 7 H) 1.62 (dd, \textit{J}=8.41, 4.11 Hz, 1 H) 1.77 (dt, \textit{J}=13.40, 3.86 Hz, 2 H) 1.92 - 2.09 (m, 7 H) 3.26 - 3.42 (m, 4 H) 3.73 - 3.88 (m, 2 H) 3.93 (t, \textit{J}=6.26 Hz, 4 H) 6.95 - 7.15 (m, 2 H) 7.44 (d, \textit{J}=9.00 Hz, 1 H) 9.65 (br. s., 1 H) 3.3-4.3 (br. s. 1H (NH)). LCMS \textit{m/z}: [M+H]\textsuperscript{+} Calcd for C \textsubscript{23}H\textsubscript{33}N\textsubscript{6}O 409.3; Found 409.0. & 1.45 \\
\midrule
A1  \textsubscript{H\textsubscript{2}} & \textsuperscript{1}H NMR (400 MHz, CDCl\textsubscript{3}): $\delta$ (ppm) 1.99 - 2.15 (m, 10 H) 2.55 (s, 3 H) 3.33 (dt, \textit{J}=17.90, 6.31 Hz, 4 H) 4.04 (br. s., 3 H) 6.98 (s, 1 H) 7.07 (s, 1 H) 7.51 (d, \textit{J}=9.00 Hz, 1 H). LCMS \textit{m/z}: [M+H]\textsuperscript{+} Calcd for C\textsubscript{18}H\textsubscript{24}N\textsubscript{5}O 326.2; Found 326.2. & >100 \\
\midrule
A2  \textsubscript{H\textsubscript{2}} & \textsuperscript{1}H NMR (600 MHz, CDCl\textsubscript{3}): $\delta$ (ppm) 1.22 - 1.35 (m, 6 H) 2.04 - 2.14 (m, 11 H) 3.38 (t, \textit{J}=6.43 Hz, 4 H) 4.12 (br. s., 3 H) 7.03 - 7.17 (m, 2 H) 7.49 (d, \textit{J}=8.77 Hz, 1 H). LCMS \textit{m/z}: [M+H]\textsuperscript{+} Calcd for C\textsubscript{20}H\textsubscript{28}N\textsubscript{5}O 354.2; Found 354.0. & 2.8 \\
\midrule
A5\textsubscript{H\textsubscript{2}} & \textsuperscript{1}H NMR (600 MHz, CDCl\textsubscript{3}): $\delta$ (ppm) 0.98 - 1.12 (m, 6 H) 2.01 - 2.17 (m, 10 H) 2.30 (dd, \textit{J}=13.15, 6.72 Hz, 1 H) 2.62 - 2.81 (m, 3 H) 3.33 (s, 1 H) 3.38 (t, \textit{J}=6.43 Hz, 2 H) 3.98 (br. s., 1 H) 4.10 (br. s., 2 H) 7.05 - 7.13 (m, 2 H) 7.48 (br. s., 1 H). LCMS \textit{m/z}: [M+H]\textsuperscript{+} Calcd for C\textsubscript{21}H\textsubscript{30}N\textsubscript{5}O 368.2; Found 368.0. & >100 \\
\midrule
B1\textsubscript{H\textsubscript{2}} & \textsuperscript{1}H NMR (400 MHz, CDCl\textsubscript{3}): $\delta$ (ppm) 1.91 - 2.04 (m, 9 H) 3.36 (br. s., 4 H) 3.89 (br. s., 4 H) 7.11 (d, \textit{J}=2.35 Hz, 1 H) 7.20 (dd, \textit{J}=9.00, 2.35 Hz, 1 H) 7.44 - 7.61 (m, 4 H) 7.88 - 8.00 (m, 2 H). LCMS \textit{m/z}: [M+H]\textsuperscript{+} Calcd for C\textsubscript{23}H\textsubscript{26}N\textsubscript{5}O 388.2; Found 388.0. & 4.35 \\
\midrule
B2\textsubscript{H\textsubscript{2}} & \textsuperscript{1}H NMR (600 MHz, CDCl\textsubscript{3}): $\delta$ (ppm) 2.00 - 2.17 (m, 9 H) 3.36 (t, \textit{J}=6.43 Hz, 4 H) 4.05 (br. s., 4 H) 7.05 - 7.12 (m, 2 H) 7.41 - 7.51 (m, 1 H) 7.99 (br. s., 2 H) 8.74 (d, \textit{J}=5.26 Hz, 2 H). LCMS \textit{m/z}: [M+H]\textsuperscript{+} Calcd for C\textsubscript{22}H\textsubscript{25}N\textsubscript{6}O 389.2; Found 389.0. & 23.6 \\
\midrule
B3\textsubscript{H\textsubscript{2}} & \textsuperscript{1}H NMR (400 MHz, CDCl\textsubscript{3}): $\delta$ (ppm) 1.76 (br. s., 6 H) 1.97 - 2.16 (m, 18 H) 3.35 (t, \textit{J}=6.46 Hz, 4 H) 4.06 (t, \textit{J}=6.46 Hz, 4 H) 6.99 - 7.13 (m, 2 H) 7.60 (d, \textit{J}=9.00 Hz, 1 H). LCMS \textit{m/z}: [M+H]\textsuperscript{+} Calcd for C\textsubscript{27}H\textsubscript{36}N\textsubscript{5}O 446.3; Found 446.0. & 1.77 \\
\midrule
B5\textsubscript{H\textsubscript{2}} & \textsuperscript{1}H NMR (400 MHz, CDCl\textsubscript{3}): $\delta$ (ppm) 1.97 - 2.15 (m, 8 H) 3.25 - 3.39 (m, 4 H) 4.11 (br. s., 3 H) 6.80 - 6.95 (m, 2 H) 7.35 - 7.46 (m, 3 H) 7.62 - 7.72 (m, 2 H) 7.83 (s, 1 H). LCMS \textit{m/z}: [M+H]\textsuperscript{+} Calcd for C\textsubscript{24}H\textsubscript{28}N\textsubscript{5}O\textsubscript{2} 418.2; Found 418.2. & 2.51 \\
\midrule
C1\textsubscript{H\textsubscript{2}} & \textsuperscript{1}H NMR (400 MHz, CDCl\textsubscript{3}): $\delta$ (ppm) 1.94 - 2.16 (m, 9 H) 3.38 (br. s., 4 H) 4.05 (br. s., 4 H) 7.07 - 7.17 (m, 2 H) 7.48 - 7.66 (m, 3 H) 7.83 - 8.02 (m, 3 H) 8.13 (d, \textit{J}=7.88 Hz, 1 H) 8.62 (br. s., 1 H). LCMS \textit{m/z}: [M+H]\textsuperscript{+} Calcd for C\textsubscript{27}H\textsubscript{28}N\textsubscript{5}O 438.2; Found 438.2. & 1.00 \\
\bottomrule
\end{tabularx}
\label{tab:hydrocompounds}
\end{table}

\begin{table}
\centering
\begin{tabularx}{\textwidth}{l|X|l}
\toprule
\textbf{Compound} & \textbf{Analytical Data} & \textbf{IC\textsubscript{50} ($\mu$M)} \\
\midrule
C2\textsubscript{H\textsubscript{2}} & \textsuperscript{1}H NMR (400 MHz, CDCl\textsubscript{3}): $\delta$ (ppm) 1.93 - 2.15 (m, 9 H) 3.36 (br. s., 4 H) 3.94 (br. s., 4 H) 4.26 (br. s., 2 H) 7.03 - 7.15 (m, 2 H) 7.26 (br. s., 1 H) 7.29 - 7.42 (m, 4 H) 7.57 (d, \textit{J}=9.01 Hz, 1 H). LCMS \textit{m/z}: [M+H]\textsuperscript{+} Calcd for C\textsubscript{24}H\textsubscript{28}N\textsubscript{5}O 402.2; Found 402.0. & 0.77 \\
\midrule
C3\textsubscript{H\textsubscript{2}} & \textsuperscript{1}H NMR (600 MHz, CDCl\textsubscript{3}): $\delta$ (ppm) 1.98 - 2.14 (m, 9 H) 3.35 - 3.43 (m, 4 H) 4.01 (t, \textit{J}=6.14 Hz, 4 H) 7.05 - 7.18 (m, 2 H) 7.34 - 7.46 (m, 3 H) 7.56 - 7.66 (m, 3 H) 7.82 (d, \textit{J}=14.62 Hz, 2 H). LCMS \textit{m/z}: [M+H]\textsuperscript{+} Calcd for C\textsubscript{25}H\textsubscript{28}N\textsubscript{5}O 414.2; Found 414.2. & 6.66 \\
\midrule
C4\textsubscript{H\textsubscript{2}} & \textsuperscript{1}H NMR (400 MHz, CDCl\textsubscript{3}): $\delta$ (ppm) 1.83 (dd, \textit{J}=12.72, 3.72 Hz, 2 H) 1.97 - 2.20 (m, 12 H) 3.04 - 3.18 (m, 2 H) 3.31 - 3.42 (m, 4 H) 4.03 (br. s., 4 H) 4.85 (d, \textit{J}=13.69 Hz, 2 H) 6.73 (d, \textit{J}=4.70 Hz, 1 H) 7.04 - 7.17 (m, 2 H) 7.51 (d, \textit{J}=9.00 Hz, 1 H) 8.49 (d, \textit{J}=4.70 Hz, 1 H). LCMS \textit{m/z}: [M+H]\textsuperscript{+} Calcd for C\textsubscript{27}H\textsubscript{32}N\textsubscript{8}OF\textsubscript{3} 541.3; Found 541.0. & 0.95 \\
\midrule
C5\textsubscript{H\textsubscript{2}} & \textsuperscript{1}H NMR (400 MHz, CDCl\textsubscript{3}): $\delta$ (ppm) 1.94 - 2.13 (m, 8 H) 3.31 (br. s., 4 H) 3.92 (br. s., 4 H) 4.67 (d, \textit{J}=5.25 Hz, 2 H) 6.92 - 7.04 (m, 2 H) 7.19 - 7.26 (m, 1 H) 7.31 (t, \textit{J}=7.13 Hz, 2 H) 7.40 (d, \textit{J}=7.75 Hz, 2 H) 7.48 (d, \textit{J}=9.01 Hz, 1 H). LCMS \textit{m/z}: [M+H]\textsuperscript{+} Calcd for C\textsubscript{23}H\textsubscript{28}N\textsubscript{5} 374.2; Found 374.0. & 2.44 \\
\midrule
D1\textsubscript{H\textsubscript{2}} & \textsuperscript{1}H NMR (400 MHz, DMSO-d\textsubscript{6}): $\delta$ (ppm) 1.96 (ddt, \textit{J}=18.54, 6.31, 3.42, 3.42 Hz, 8 H) 2.91 (s, 6 H) 3.27 - 3.31 (m, 4 H) 3.87 (t, \textit{J}=6.26 Hz, 4 H) 7.06 (d, \textit{J}=2.35 Hz, 1 H) 7.12 (d, \textit{J}=9.00 Hz, 1 H) 7.40 (d, \textit{J}=9.00 Hz, 1 H). LCMS \textit{m/z}: [M+H]\textsuperscript{+} Calcd for C\textsubscript{19}H\textsubscript{26}N\textsubscript{6}O 355.2; Found 355.0. & >100 \\ 
\midrule
D2\textsubscript{H\textsubscript{2}} & \textsuperscript{1}H NMR (400 MHz, CDCl\textsubscript{3}): $\delta$ (ppm) 1.22 - 1.36 (m, 6 H) 1.95 - 2.18 (m, 8 H) 3.27 - 3.43 (m, 4 H) 3.91 (br. s., 4 H) 4.09 (dd, \textit{J}=13.11, 6.46 Hz, 1 H) 6.89 (br. s., 1 H) 7.00 - 7.12 (m, 2 H) 7.44 (d, \textit{J}=9.00 Hz, 1 H) 9.56 (br. s., 1 H). LCMS \textit{m/z}: [M+H]\textsuperscript{+} Calcd for C\textsubscript{20}H\textsubscript{29}N\textsubscript{6}O 369.2; Found 369.2. & 5.66 \\
\midrule
D3\textsubscript{H\textsubscript{2}} & \textsuperscript{1}H NMR (400 MHz, DMSO-d\textsubscript{6}+TFA additive): $\delta$ (ppm) 1.82 - 2.11 (m, 8 H) 3.15 - 3.53 (m, 11 H) 3.80 (br. s., 2 H) 4.28 (br. s., 2 H) 7.14 (d, \textit{J}=2.35 Hz, 1 H) 7.19 - 7.31 (m, 2 H) 7.85 (d, \textit{J}=9.39 Hz, 1 H) 10.40 (s, 1 H). LCMS \textit{m/z}: [M+H]\textsuperscript{+} Calcd for C\textsubscript{20}H\textsubscript{29}N\textsubscript{6}O\textsubscript{2} 385.2; Found 385.2. & >100 \\
\midrule
D4\textsubscript{H\textsubscript{2}} & \textsuperscript{1}H NMR (400 MHz, CDCl\textsubscript{3}): $\delta$ (ppm) 2.00 - 2.11 (m, 8 H) 3.37 (t, \textit{J}=6.46 Hz, 4 H) 3.96 (t, \textit{J}=6.46 Hz, 4 H) 6.98 - 7.15 (m, 4 H) 7.31 - 7.38 (m, 2 H) 7.54 (d, \textit{J}=9.00 Hz, 1 H) 7.66 (d, \textit{J}=7.83 Hz, 2 H) 12.23 (br. s., 1 H). LCMS \textit{m/z}: [M+H]\textsuperscript{+} Calcd for C\textsubscript{23}H\textsubscript{27}N\textsubscript{6}O 403.2; Found 403.0. & 6.00 \\
\midrule
D5\textsubscript{H\textsubscript{2}} & \textsuperscript{1}H NMR (400 MHz, CDCl\textsubscript{3}): $\delta$ (ppm) 2.01 - 2.16 (m, 8 H) 3.31 - 3.42 (m, 4 H) 3.91 - 4.03 (m, 7 H) 7.01 - 7.17 (m, 3 H) 7.42 (t, \textit{J}=8.02 Hz, 1 H) 7.57 (d, \textit{J}=9.00 Hz, 1 H) 7.74 (d, \textit{J}=7.83 Hz, 1 H) 8.04 (d, \textit{J}=7.83 Hz, 1 H) 8.17 (s, 1 H) 12.41 (br. s., 1 H). LCMS \textit{m/z}: [M+H]\textsuperscript{+} Calcd for C\textsubscript{25}H\textsubscript{29}N\textsubscript{6}O\textsubscript{3} 461.2; Found 461.2. & >100 \\
\bottomrule
\end{tabularx}
\end{table}

\begin{table}
\centering
\begin{tabularx}{\textwidth}{l|X|l}
\toprule
\textbf{Compound} & \textbf{Analytical Data} & \textbf{IC\textsubscript{50} ($\mu$M)} \\
\midrule
E1\textsubscript{H\textsubscript{2}} & \textsuperscript{1}H NMR (400 MHz, CDCl\textsubscript{3}): $\delta$ (ppm) 1.87 - 2.12 (m, 8 H) 3.33 (t, \textit{J}=6.46 Hz, 4 H) 3.78 (br. s., 4 H) 4.62 (d, \textit{J}=5.48 Hz, 2 H) 6.94 - 7.10 (m, 3 H) 7.29 (d, \textit{J}=7.43 Hz, 1 H) 7.31 - 7.48 (m, 5 H) 10.05 (br. s., 1 H). LCMS \textit{m/z}: [M+H]\textsuperscript{+} Calcd for C\textsubscript{24}H\textsubscript{29}N\textsubscript{6}O 417.2; Found 417.0. & 1.15 \\
\midrule
E2\textsubscript{H\textsubscript{2}} & \textsuperscript{1}H NMR (600 MHz, CDCl\textsubscript{3}): $\delta$ (ppm) 1.51 - 1.80 (m, 12 H) 1.93 - 2.17 (m, 11 H) 3.11 (d, \textit{J}=5.26 Hz, 2 H) 3.29 - 3.42 (m, 4 H) 3.91 (br. s., 4 H) 6.95 (br. s., 1 H) 7.03 - 7.14 (m, 2 H) 7.45 (d, \textit{J}=9.35 Hz, 1 H) 9.89 (br. s., 1 H). LCMS \textit{m/z}: [M+H]\textsuperscript{+} Calcd for C\textsubscript{28}H\textsubscript{39}N\textsubscript{6}O 475.3; Found 475.2. & 2.45 \\
\midrule
E3\textsubscript{H\textsubscript{2}} & \textsuperscript{1}H NMR (400 MHz, CDCl\textsubscript{3}): $\delta$ (ppm) 1.92 - 2.12 (m, 8 H) 2.92 (s, 3 H) 3.36 (br. s., 4 H) 3.44 (t, \textit{J}=7.82 Hz, 2 H) 4.02 (br. s., 4 H) 4.10 (t, \textit{J}=7.88 Hz, 2 H) 7.04 - 7.17 (m, 2 H) 7.59 (d, \textit{J}=8.88 Hz, 1 H). LCMS \textit{m/z}: [M+H]\textsuperscript{+} Calcd for C\textsubscript{20}H\textsubscript{27}N\textsubscript{6}O 367.2; Found 367.0. & 90.03 \\
\midrule
E4\textsubscript{H\textsubscript{2}} & \textsuperscript{1}H NMR (400 MHz, CDCl\textsubscript{3}): $\delta$ (ppm) 2.06 (d, \textit{J}=6.50 Hz, 8 H) 3.37 (br. s., 4 H) 3.96 (t, \textit{J}=7.75 Hz, 2 H) 4.05 (br. s., 4 H) 4.19 - 4.35 (m, 2 H) 7.03 - 7.17 (m, 3 H) 7.38 (t, \textit{J}=7.50 Hz, 2 H) 7.56 - 7.72 (m, 3 H). LCMS \textit{m/z}: [M+H]\textsuperscript{+} Calcd for C\textsubscript{25}H\textsubscript{29}N\textsubscript{6}O 429.2; Found 429.0. & 1.42 \\
\bottomrule
\end{tabularx}
\end{table}


\begin{table}[htbp]
\centering
\caption{Compound Characterization Data of \textit{Monofluorinated} Quinazoline Compounds (8-fluoro-4,6-di(pyrrolidin-1-yl)quinazoline). Reference to Figure~\ref{fig:synthesized_mols} for the corresponding compound, where $X = H, F$} 
\begin{tabularx}{\textwidth}{l|X|l}
\toprule
\textbf{Compound} & \textbf{Analytical Data} & \textbf{IC\textsubscript{50} ($\mu$M)} \\
\midrule

A1\textsubscript{H, F} & \textsuperscript{1}H NMR (400   MHz, CDCl\textsubscript{3}): $\delta$ (ppm) 2.01 - 2.17   (m, 8 H) 2.60 (s, 3 H) 3.35 (t, \textit{J}=6.46 Hz, 4 H) 4.16 (br. s., 4 H) 6.83 -   6.92 (m, 2 H). LCMS \textit{m/z}: [M+H]\textsuperscript{+} Calcd for C\textsubscript{18}H\textsubscript{23}N\textsubscript{5}OF   344.2; Found 344.0. &
  \textgreater{}100 \\
\midrule
A2\textsubscript{H, F} & \textsuperscript{1}H NMR (600   MHz, CDCl\textsubscript{3}): $\delta$ (ppm) 1.28 (d,   \textit{J}=7.02 Hz, 6 H) 2.00 - 2.12 (m, 8 H) 3.31 - 3.38 (m, 4 H) 4.01 (br. s., 4 H)   6.83 - 6.89 (m, 2 H). LCMS \textit{m/z}: [M+H]\textsuperscript{+} Calcd for C\textsubscript{20}H\textsubscript{27}N\textsubscript{5}OF   372.2; Found 372.0. &
  \textgreater{}100 \\
\midrule
A3\textsubscript{H, F} & \textsuperscript{1}H NMR (400   MHz, CDCl\textsubscript{3}): $\delta$ (ppm) 1.34 (s, 9 H)   1.96 - 2.09 (m, 8 H) 3.31 (t, \textit{J}=6.46 Hz, 4 H) 4.02 (t, \textit{J}=6.65 Hz, 4 H) 6.79 -   6.88 (m, 2 H). LCMS \textit{m/z}: [M+H]\textsuperscript{+} Calcd for C\textsubscript{21}H\textsubscript{29}N\textsubscript{5}OF   386.2; Found 386.0. &
  \textgreater{}100 \\
\midrule
A4\textsubscript{H, F} & \textsuperscript{1}H NMR (400   MHz, CDCl\textsubscript{3}): $\delta$ (ppm) 1.02 (t, \textit{J}=7.43   Hz, 3 H) 1.78 (dt, \textit{J}=1.00 Hz, 2 H) 1.99 - 2.16 (m, 9 H) 2.89 (br. s., 2 H)   3.29 - 3.39 (m, 4 H) 4.02 (br. s., 4 H) 6.81 - 6.90 (m, 2 H). LCMS \textit{m/z}:   [M+H]\textsuperscript{+} Calcd for C\textsubscript{20}H\textsubscript{27}N\textsubscript{5}OF   372.2; Found 372.0. &
  \textgreater{}100 \\
\midrule
A5\textsubscript{H, F} & \textsuperscript{1}H NMR (400   MHz, CDCl\textsubscript{3}): $\delta$ (ppm) 0.96 - 1.06   (m, 7 H) 1.99 - 2.10 (m, 8 H) 2.21 - 2.32 (m, 1 H) 2.76 (br. s., 2 H) 3.32   (t, \textit{J}=6.46 Hz, 4 H) 3.98 (br. s., 4 H) 6.78 - 6.87 (m, 2 H). LCMS \textit{m/z}:   [M+H]\textsuperscript{+} Calcd for C\textsubscript{21}H\textsubscript{29}N\textsubscript{5}OF   386.2; Found 386.0. &
  \textgreater{}100 \\
\midrule
B1\textsubscript{H, F} & \textsuperscript{1}H NMR (400   MHz, DMSO-d\textsubscript{6}): $\delta$ (ppm) 1.84 - 2.05   (m, 8 H) 3.34 (br. s., 4 H(hidden)) 3.84 (br. s., 4 H) 6.88 - 6.96 (m, 1 H)   7.05 (dd, \textit{J}=13.69, 2.35 Hz, 1 H) 7.39 - 7.52 (m, 2 H) 7.52 - 7.60 (m, 1 H)   7.86 - 7.98 (m, 2 H) 10.38 (s, 1 H). LCMS \textit{m/z}: [M+H]\textsuperscript{+} Calcd   for C\textsubscript{23}H\textsubscript{25}N\textsubscript{5}OF 406.2; Found 406.0. &
  14.55 \\
\midrule
B4\textsubscript{H, F} & \textsuperscript{1}H NMR (600   MHz, CDCl\textsubscript{3}): $\delta$ (ppm) 1.96 - 2.11   (m, 8 H) 3.24 - 3.37 (m, 4 H) 3.48 (s, 1 H) 3.86 (s, 3 H) 4.02 (t, \textit{J}=6.46 Hz,   4 H) 6.79 - 6.89 (m, 2 H) 6.95 (m, \textit{J}=8.61 Hz, 2 H) 7.98 (m, \textit{J}=8.61 Hz, 2 H).   LCMS \textit{m/z}: [M+H]\textsuperscript{+} Calcd for C\textsubscript{24}H\textsubscript{27}N\textsubscript{25}O\textsubscript{2}F   436.2; Found 436.0. &
  6.66 \\
\midrule
C3\textsubscript{H, F} & \textsuperscript{1}H NMR (400   MHz, CDCl\textsubscript{3}): $\delta$ (ppm) 1.62 (br. s.,   1 H) 1.98 - 2.19 (m, 8 H) 3.25 - 3.39 (m, 4 H) 4.12 (br. s., 4 H) 6.82 - 6.94   (m, 2 H) 7.33 - 7.45 (m, 3 H) 7.62 - 7.73 (m, 2 H) 7.75 - 7.97 (m, 2 H). LCMS   \textit{m/z}: [M+H]\textsuperscript{+} Calcd for C\textsubscript{25}H\textsubscript{27}N\textsubscript{5}OF   432.2; Found 432.0. &
  8.73 \\

\bottomrule
\end{tabularx}
\label{tab:monofluorocompounds}
\end{table}

\begin{table}[htbp]
\centering
\caption{Compound Characterization Data of \textit{Difluorinated} Quinazoline Compounds (7, 8-fluoro-4,6-di(pyrrolidin-1-yl)quinazoline). Reference to Figure~\ref{fig:synthesized_mols} for the corresponding compound, where $X = F$} 
\begin{tabularx}{\textwidth}{l|X|l}
\toprule
\textbf{Compound} & \textbf{Analytical Data} & \textbf{IC\textsubscript{50} ($\mu$M)} \\
\midrule

A5\textsubscript{F\textsubscript{2}} & \textsuperscript{1}H NMR (400   MHz, CDCl\textsubscript{3}): $\delta$ (ppm) 1.03 (d,   \textit{J}=6.65 Hz, 6 H) 1.61 (br. s., 1 H(hidden)) 1.95 - 2.13 (m, 8 H) 2.20 - 2.37   (m, 1 H) 2.76 (br. s., 2 H) 3.46 (br. s., 4 H) 3.96 (br. s., 4 H) 7.02 (d,   \textit{J}=9.00 Hz, 1 H). LCMS \textit{m/z}: [M+H]\textsuperscript{+} Calcd for C\textsubscript{21}H\textsubscript{28}N\textsubscript{5}OF\textsubscript{2} 404.2;   Found 404.0. &
  2.32 \\
\midrule
B1\textsubscript{F\textsubscript{2}} & \textsuperscript{1}H NMR (600   MHz, CDCl\textsubscript{3}): $\delta$ (ppm) 0.08 (s, 1 H)   1.93 - 2.20 (m, 8 H) 3.39 - 3.54 (m, 4 H) 3.91 - 4.07 (m, 4 H) 7.00 - 7.11   (m, 1 H) 7.43 - 7.61 (m, 3 H) 7.98 (d, \textit{J}=7.43 Hz, 2 H). LCMS \textit{m/z}:   [M+H]\textsuperscript{+} Calcd for C\textsubscript{23}H\textsubscript{24}N\textsubscript{5}OF\textsubscript{2} 424.2;   Found 424.0. &
  3.39 \\
\midrule
C3\textsubscript{F\textsubscript{2}} & \textsuperscript{1}H NMR (400   MHz, CDCl\textsubscript{3}): $\delta$ (ppm) 1.84 - 2.34   (m, 8 H) 3.48 (br. s., 4 H) 4.07 (br. s., 4 H) 6.98 - 7.13 (m, 1 H) 7.39 (br.   s., 3 H) 7.64 (br. s., 2 H) 7.87 (br. s., 1 H). LCMS \textit{m/z}: [M+H]\textsuperscript{+}   Calcd for C\textsubscript{25}H\textsubscript{26}N\textsubscript{5}OF\textsubscript{2} 450.2;   Found 450.2. &
  1.48 \\
\midrule
C4\textsubscript{F\textsubscript{2}} & \textsuperscript{1}H NMR (400   MHz, CDCl\textsubscript{3}): $\delta$ (ppm) 1.31 - 1.91   (m, 8 H) 1.91 - 2.10 (m, 10 H) 2.13 (br. s., 1 H) 3.10 (t, \textit{J}=11.93 Hz, 2 H)   3.47 (br. s., 4 H) 3.91 (br. s., 1 H) 4.01 (br. s., 4 H) 4.85 (d, \textit{J}=12.91 Hz,   2 H) 6.72 (d, \textit{J}=4.70 Hz, 1 H) 7.00 (d, \textit{J}=7.83 Hz, 1 H) 8.47 (d, \textit{J}=5.09 Hz, 1   H). LCMS \textit{m/z}: [M+H]\textsuperscript{+} Calcd for C\textsubscript{27}H\textsubscript{30}N\textsubscript{8}OF\textsubscript{5} 577.2;   Found 577.2. &
  1.28 \\
\midrule
E1\textsubscript{F\textsubscript{2}} & \textsuperscript{1}H NMR (600   MHz, CDCl\textsubscript{3}): $\delta$ (ppm) 1.91 - 2.09   (m, 8 H) 3.35 - 3.45 (m, 4 H) 3.77 (br. s., 4 H) 4.61 (d, \textit{J}=5.26 Hz, 2 H)   6.92 - 7.03 (m, 1 H) 7.08 (br. s., 1 H) 7.28 (s, 1 H) 7.36 (t, \textit{J}=7.60 Hz, 2   H) 7.44 (d, \textit{J}=7.60 Hz, 2 H) 10.01 (br. s., 1 H). LCMS \textit{m/z}: [M+H]\textsuperscript{+}   Calcd for C\textsubscript{24}H\textsubscript{27}N\textsubscript{6}OF\textsubscript{2} 453.2;   Found 453.2. &
  4.34 \\

\bottomrule
\end{tabularx}
\label{tab:difluorocompounds}
\end{table}
\end{document}